\documentclass[showpacs,preprintnumbers,amsmath,amssymb]{revtex4}
\usepackage[pdftex]{epsfig}
\DeclareGraphicsExtensions{.jpg,.mps,.pdf,.png}
\usepackage{dcolumn}% Align table columns on decimal point
\usepackage{bm}% bold math

\newcommand{\hq}{\hat{q}}

\newcommand{\vq}{{\bf q}}

\newcommand{\vk}{{\bf k}}

\newcommand{\dd}{{\rm d}}

\newcommand{\mg}{\big<}
\newcommand{\md}{\big>}

\newcommand{\mF}{{\cal F}}
\newcommand{\mH}{{\cal H}}

\newcommand{\mW}{{\cal W}}

\newcommand{\Dirac}{\delta_{\rm D}}

\newcommand{\beq}{\begin{equation}}
\newcommand{\eeq}{\end{equation}}
\newcommand{\beqa}{\begin{eqnarray}}
\newcommand{\eeqa}{\end{eqnarray}}

\def\fun#1#2{\lower3.6pt\vbox{\baselineskip0pt\lineskip.9pt
        \ialign{$\mathsurround=0pt#1\hfill##\hfil$\crcr#2\crcr\sim\crcr}}}

\def\Mpc{\, h^{-1} \, {\rm Mpc}}

\newcommand{\bdm}{\begin{displaymath}}
\newcommand{\edm}{\end{displaymath}}
\newcommand{\bea}{\begin{eqnarray}}
\newcommand{\eea}{\end{eqnarray}}
\newcommand{\bt}{\begin{tabular}}
\newcommand{\et}{\end{tabular}}

\newcommand{\oneloop}{{\rm 1-loop}}
\newcommand{\twoloops}{{\rm 2-loop}}
\newcommand{\threeloops}{{\rm 3-loop}}
\newcommand{\ploops}{{\rm p-loop}}
\newcommand{\nloops}{{\rm n-loop}}

\newcommand{\ct}{{\rm c.t.}}
\newcommand{\vf}{{\cal F}}
\newcommand{\vg}{{\cal G}}
\newcommand{\RPT}{{\rm RPT}}
\newcommand{\reg}{{\rm reg}}
\newcommand{\sigmad}{\sigma_{\rm d}}

\newcommand{\sigmadlin}{\sigma_{\rm d, lin.}}

\newcommand{\RegG}{{^{\rm Reg}G}}

\def\Mpc{{\rm Mpc}}

\def\ltsima{$\; \buildrel < \over \sim \;$}   
\def\gtsima{$\; \buildrel > \over \sim \;$} 
\def\simlt{\lower.5ex\hbox{\ltsima}}   
\def\simgt{\lower.5ex\hbox{\gtsima}}

\begin{document}
%\title{Cosmic propagators at two and three-loop orders}
\title{Cosmic propagators at two-loop order}

\author{Francis Bernardeau}
\affiliation{Institut de Physique ThŽorique,
CEA, IPhT, F-91191 Gif-sur-Yvette, France}
\affiliation{CNRS, URA 2306, F-91191 Gif-sur-Yvette, France}
\author{Atsushi Taruya}
\affiliation{Research Center for the Early Universe, School of Science, 
The University of Tokyo, Bunkyo-ku, Tokyo 113-0033, Japan}
\affiliation{Kavli Institute for the Physics and Mathematics of the Universe,
Todai Institutes for Advanced Study, the University of Tokyo,
Kashiwa, Chiba 277-8583, Japan (Kavli IPMU, WPI)}
\author{Takahiro Nishimichi}
\affiliation{Kavli Institute for the Physics and Mathematics of the Universe,
Todai Institutes for Advanced Study, the University of Tokyo,
Kashiwa, Chiba 277-8583, Japan (Kavli IPMU, WPI)}
%\affiliation{Institute for the Physics and Mathematics of the Universe,
%University of Tokyo, Kashiwa, Chiba 277-8568, Japan}
\vspace{.2 cm}
\date{\today}
\vspace{.2 cm}
\begin{abstract}
We  explore the properties of two-point cosmic propagators when Perturbation Theory (PT)
loop corrections are consistently taken into account. We show in particular how the interpolation scheme 
proposed in \cite{2011arXiv1112.3895B} can be explicitly used up to two-loop order introducing the notion
of regular parts for the contributing terms. Extending the one-loop results, we then derive and give semi analytical forms of the two-loop contributions for both the cosmic density and velocity propagators. These results are tested against numerical simulations and shown to significantly improve upon 
one-loop results for redshifts above $0.5$. \\
We found however that  at lower redshift  two-loop order corrections are  too large partly due to a strong sensitivity of those terms to the 
small scale modes. We show that this dependence is expected to be even larger for higher order loop corrections 
both from theoretical investigations and numerical tests, the latter obtained with Monte Carlo evaluations of the three-loop
contributions. This  makes small-scale regularization schemes necessary for the use of such higher order corrections.
\end{abstract}
\vskip2pc

\maketitle

\section{Introduction}

Precision measurements  of the statistical properties of  the large-scale structure  of the universe is the key to current large projects of observational cosmology. They aim in particular at measuring the galaxy power spectrum at the scales where Baryonic Acoustic Oscillations can be detected 
(\cite{2005ApJ...633..560E,2007ApJ...657...51P,2011MNRAS.415.2892B,2012arXiv1202.6057N}) or at measuring the projected density power spectrum from weak lensing observations (as stressed in \cite{2006astro.ph..9591A} and which is the part of the core program of the Euclid mission \cite{2011arXiv1110.3193L}). In this context,
semi-analytic tools are now developed that aim at computing, from first principles,  the statistical properties of the cosmic fields  at large-scale. Standard Perturbation Theory approaches (see \cite{2002PhR...367....1B}) are however not very efficient in producing well behaved next-to-leading order terms to the spectra or bispectra. It leads indeed to a resummation series that has poor converging properties (see \cite{2006PhRvD..73f3519C} for more insights into that). Alternative approaches have been recently developed that propose, implicitly or explicitly, a reorganization of the perturbation series. This is the case in particular for the  Renormalization Perturbation Theory (RPT) approaches developed initially in \cite{2006PhRvD..73f3519C}. Other approaches are based on the writing of closed system of integro-differential equations as in the closure theory approach, \cite{2008ApJ...674..617T}, or in the Time Flow equation hierarchy presented in \cite{2008JCAP...10..036P}. 
We put ourselves in the context of the RPT  approaches and its extensions which are based on the introduction of the two-point propagator and on a re-writing of the series expansion with the help of that object. A related but alternative approach has been proposed in \cite{2008PhRvD..78j3521B} where was introduced the concepts of multi-point propagators. Interaction vertices can then be formally renormalized as well. It has been shown there that a new scheme for the computation of spectra and poly-spectra can be obtained form the multi-point propagators, from the the so-called $\Gamma$ expansion.  
% therefore seen as the building blocks of the Perturbation Theory calculations. 

The focus of this paper is precisely on the behavior of the two-point propagator and on the precision one can theoretically attain in its description. In particular one wishes to evaluate the importance of the multi-loop corrections to its expression, in practice up to second order, and to see whether these corrections can be safely computed. Comparisons with N-body codes will be used taking advantage of the interpolation scheme proposed in \cite{2011arXiv1112.3895B}. In this paper it is indeed shown how multi-loop corrections can be consistently used to get predictions on multi-propagator expressions with theoretically increasing precision.
These investigations aim in particular at pinning down the validity range of this type of calculations. As we will see in the following the 
two-point propagator is the quantity, at a given order in PT, that is the most sensitive to high-$k$ modes in the computation of power spectra. 
Quantifying the importance of the small scale physics, for large-scale predictions, is a daunting quest. Some phenomenological results have been proposed in \cite{2009PhRvD..80d3504P,2010arXiv1009.0106V} to investigate for instance the importance of shell crossings. This is also the idea behind the works in \cite{PieManSav1108,2012arXiv1206.2926C} where the impact of the small scale physics is tentatively introduced as extra source terms. In this paper we eventually investigate this problem from a PT approach.

The paper is organized as follows. In Sec.~\ref{sec:motionequations}
we recall the motion equations for gravitational instabilities of a cosmic fluid. In Sec.~\ref{sec:thediagrams} we 
compute the two-loop corrective terms and propose analytic fits for them. 
In Sect.  \ref{sec:multiloops}  we first estimate the contributions of the three-loop terms from Monte-Carlo computations.
Our results are tested with the help of consistency relation we provided in Sect. \ref{sec:thediagrams}.
We then explore in more detail the 
expected properties of the loop corrections at arbitrary orders introducing the notion of kernels and henceforth quantifying the sensitivity of 
the results to large and small scales modes. Our concluding remarks are presented in Sec.~\ref{sec:conclusions}.

\section{The cosmic  two-point propagators}
\label{sec:motionequations}

\subsection{The equations of motion}

We are interested here in the development of cosmological instabilities in a cosmological dust fluid. In general the dynamical evolution of such a fluid can be described with the
Vlasov equation. As usual we then restrict our investigations to the regime where multi-flow regions play a negligible role but we will discuss this assumption later on. In the one flow limit, the motion equation then takes the
form of a set of three coupled equations relating the local density contrast, the peculiar
velocity field and the gravitational potential (see \cite{2002PhR...367....1B}). 

At {\it linear order} these equations can easily be solved for an arbitrary background cosmology.
One generically finds a growing solution and a decaying solution. Let us denotes $D_{+}(\tau)$ the growing 
mode solution for the density contrast $f_{+}(\tau)$ its logarithmic derivative with respect to the expansion
so that,
\begin{equation}
\delta(\vk,\eta)=D_{+}(\eta)\delta_{0}(\vk),\ \ \theta(\vk,\eta)/\mH=-f_{+}(\eta)D_{+}(\eta)\delta_{0}(\vk)
\end{equation}
is the solution for the growing mode and similarly,
\begin{equation}
\delta(\vk,\eta)=D_{-}(\eta)\delta_{0}(\vk),\ \ \theta(\vk,\eta)/\mH=-f_{-}(\eta)D_{-}(\eta)\delta_{0}(\vk)
\end{equation}
for the decaying.

Following \cite{1998MNRAS.299.1097S}, the motion equations describing a pressure-less fluid in the one-flow limit can be written
in a compact form with the use of the doublet  $\Psi_a(\vk,\tau)$, defined as,
\begin{equation}
\Psi_a(\vk,\tau) \equiv \Big( \delta(\vk,\tau),\ -\frac{1}{f_{+}(\tau)\cal H}\theta(\vk,\tau) \Big),
\label{2vector}
\end{equation}
where
${\cal H}\equiv {d\ln a
/{d\tau}}$  is the conformal expansion rate  with  $a(\tau)$ the cosmological scale factor and $\tau$ the conformal time and where the 
index $a=1,2$ selects the density or velocity components and which makes explicit use of the growing solution.

It is then convenient to re-express the time dependence in terms of the growing solution and in the following we will use the time variable $\eta$
defined as 
\begin{equation}
\eta=\log {D_{+}(\eta)}
\end{equation}
assuming the growing factor is set to unity at initial time. Then the {\it fully nonlinear} equations of
motion in Fourier space read \cite{2002PhR...367....1B} (we henceforth use the convention that  repeated
Fourier arguments are integrated over and the Einstein convention on repeated indices),
\begin{eqnarray}
\frac{\partial}{\partial \eta} \Psi_a(\vk,\eta) + \Omega_{ab}(\eta) \Psi_b(\vk,\eta) &=& \gamma_{abc}(\vk,\vk_1,\vk_2) \ \Psi_b(\vk_1,\eta) \ \Psi_c(\vk_2,\eta),
\label{eom}
\end{eqnarray}
where 
\begin{equation}
\Omega_{ab} (\eta) \equiv \Bigg[ 
\begin{array}{cc}
0 & -1 \\ -\frac{3}{2}\frac{\Omega_{m}}{f_{+}^2} & \frac{3}{2}\frac{\Omega_{m}}{f_{+}^2}-1 
\end{array}        \Bigg],
\end{equation}
and the {\sl symmetrized vertex} matrix $\gamma_{abc}$ describes the non linear 
interactions between different Fourier modes. Its components are given by
\begin{eqnarray}
\gamma_{222}(\vk,\vk_1,\vk_2)&=&\Dirac(\vk-\vk_1-\vk_2) \ {|\vk_1+\vk_2|^2 (\vk_1
\cdot\vk_2 )\over{2 k_1^2 k_2^2}}, \nonumber \\
\gamma_{121}(\vk,\vk_1,\vk_2)&=&\Dirac(\vk-\vk_1-\vk_2) \  {(\vk_1+\vk_2) \cdot
\vk_1\over{2 k_1^2}},
\label{vertexdefinition}
\end{eqnarray}
$\gamma_{abc}(\vk,\vk_a,\vk_b)=\gamma_{acb}(\vk,\vk_b,\vk_a)$, and $\gamma=0$ 
otherwise, where $\Dirac$ denotes the Dirac delta distribution. The matrix  $\gamma_{abc}$ is independent 
on time (and on the background evolution) and encodes all the
non-linear couplings of the system.
 The formal integral solution to Eq.
(\ref{eom}) is given by (see \cite{1998MNRAS.299.1097S,2001NYASA.927...13S,2006PhRvD..73f3519C} for a detailed derivation) 
\begin{eqnarray}
\Psi_a(\vk,\eta) &=& g_{ab}(\eta) \ \phi_b(\vk) +  \int_0^{\eta}  {\dd \eta'} \ g_{ab}(\eta,\eta') \ \gamma_{bcd}^{(\rm s)}(\vk,\vk_1,\vk_2) \Psi_c(\vk_1,\eta') \Psi_d(\vk_2,\eta'),
\label{eomi}
\end{eqnarray}
where  $\phi_a(\vk)\equiv\Psi_a(\vk,\eta=0)$ denotes the initial conditions, set when the growth factor $D_{+}=1$ and where $g_{ab}(\eta)$ is the {\em linear propagator} that is Green's
function of the linearized version of Eq.~(\ref{eom}) and describes the standard linear evolution of
the density and velocity fields from their initial state. 

In the following calculations we will be using the value of the $\Omega_{ab}$ matrix to be that of the Einstein de Sitter background
that is effectively assuming that $D_{-}$ scales like $D_{+}^{-3/2}$. This is known to be a very good approximation even in the context of a $\Lambda-$CDM universe (see for instance \cite{2012arXiv1207.1465C} for an explicit investigation of the consequences of such an approximation).

The ensemble average of any
quantity can then be built out of those of the initial fields. They are 
entirely defined from the initial power spectrum of density fluctuations $P_{ab}(k)$ define as,
\begin{equation}
\mg\phi_{a}(\vk)\phi_{b}(\vk')\md=\Dirac(\vk+\vk')P_{ab}(k).
\label{Spectra}
\end{equation}  
In what follows most of the calculations and applications will be made assuming initial conditions in
the growing mode, for which $\phi_a(\vk) = \delta_0(\vk) u_a$ with $u_{a}=(1,1)$, and therefore with $P_{ab}(k)=P_0(k) u_a u_b$.

\subsection{Definition and general properties of the two-point and multi-point propagators}

The $g_{ab}(s)$  appears clearly to be one of the key ingredient of these constructions. It can be
described as the the linear propagator giving the variation of the mode amplitude as time is running.
The idea at the heart of the RPT approach is to generalize this operator beyond linear
theory. More specifically the quantity $\frac{\partial \Psi_{a}(\vk,s)}{ \partial \phi_{b}(\vk')}$
expresses
the way $\Psi_{a}(\vk,s)$  depends on $\phi_{b}(\vk')$ as a function of time $s$. This function
however depends on the stochastic properties of the fields. One can nonetheless define its
ensemble average, $G_{ab}(\vk)$, as
\begin{equation}
\left<\frac{\partial \Psi_{a}(\vk,\eta)}{\partial
\phi_{b}(\vk')}\right>=\Dirac{\left(\vk-\vk'\right)}\,G_{ab}(k;\eta).
\label{Gabdef}
\end{equation} 
This quantity depends on the initial fluctuations through the mode couplings. The ensemble 
average is made precisely over these modes. The Dirac-$\delta$ function it leads to is due, as
usual, to the homogeneity of the underlying statistical process. More generally one can define 
the $(p+1)$-point propagators $\Gamma^{(p)}$ as (see \cite{2008PhRvD..78j3521B}),
\begin{equation}
\frac{1}{p!}\left<
\frac{\delta^{p}\Psi_{a}(\vk,\eta)}{\delta\phi_{b_{1}}(\vk_{1})\dots\delta\phi_{b_{p}}(\vk_{p})}
\right>=\Dirac{\left(\vk-\vk_{1}-\dots -\vk_{p}\right)}\,
\Gamma^{(p)}_{ab_{1}\dots b_{p}}(\vk_{1},\dots,\vk_{p};\eta).
\label{Gammapdef}
\end{equation}
The Dirac $\delta$-function that appears in this expression is a direct consequence of the assumed statistical isotropy of space. 
Note that the propagators can be defined for any time interval. We will however use it here assuming the early time to be the initial time for which one can assume that the cosmic modes correspond to the Matter Dominated era growing mode (see \cite{2011JCAP...10..037A} for 
explicit investigations of the validity of such an approximation). In computing the power spectrum or bispectrum, it is convenient to introduce the reduced propagators as
\begin{equation}
\Gamma_{a}^{(p)}(\vk_{1},\dots,\vk_{p};\eta)=\Gamma^{(p)}_{ab_{1}\dots b_{p}}(\vk_{1},\dots,\vk_{p};\eta) u_{b_{1}}\dots u_{b_{p}}.
\end{equation}

Our main interest here is the late time evolution of the propagators keeping only the most growing modes contributions. 
In particular, in the following we compute the expression of $G_{ab}$ order by order in perturbation theory, up to two loop order.
Such results can be obtained from a formal expansion of $\Psi_{a}(\vk,\eta)$ with respect to the initial field,
\begin{equation}
\Psi_{a}(\vk,\eta)=\sum_{n=1}^{\infty}\Psi_{a}^{(n)}(\vk,\eta)
\label{PsiExpansion}
\end{equation}
with
%\begin{widetext}
\begin{eqnarray}
\Psi_{a}^{(n)}(\vk,\eta)=\int\dd^3\vk_{1}\dots\dd^3\vk_{n}\ 
\Dirac(\vk-\vk_{1\dots n})
% \nonumber \\
%\nonumber\\&&\hspace{-1cm}\times 
%\times 
\,\mF^{(n)}_{a b_1 b_2 \ldots b_n}(\vk_{1},\dots,\vk_{n};\eta)
\phi_{b_1}(\vk_{1})\dots\phi_{b_n}(\vk_{n})
\label{mFndef}
\end{eqnarray}
where $\mF^{(n)}$ are fully symmetric functions of the wave-vectors.  Note that these functions have
in general a non-trivial time dependence because they also include sub-leading terms in $\eta$. 
Their fastest growing term is of course given by the well known $\{F_n,G_n\}$ kernels in PT (assuming
growing mode initial conditions),
\begin{equation}
\mF^{(n)}_{a b_1 b_2 \ldots b_n}(\vk_{1},\dots,\vk_{n};\eta)u_{b_{1}}\ldots u_{b_{n}}=\exp(n\eta)\ F_a^{(n)}(\vk_1,..,\vk_n)
\nonumber
\end{equation}
with $F_{1}^{(n)}=F_{n}$ and $F_{2}^{(n)}=G_{n}$ (density or velocity divergence fields respectively).
Those quantities can obviously be computed in a standard Perturbation Theory approach. Formally one can then writes,
\begin{equation}
\Gamma_{a}^{(p)}=\Gamma_{a,\rm tree}^{(p)}+\Gamma_{a,\oneloop}^{(p)}+\Gamma_{a,\twoloops}^{(p)}+\Gamma_{a,\threeloops}^{(p)}+\dots.
\end{equation}
When keeping, for each term, the only fastest growing mode, the expressions for the perturbative corrections, $\Gamma_{a,\nloops}^{(p)}$
can be rather simplified and written in terms of the standard PT kernels. Those expressions can be written more specifically for
the two-point propagators.  Let us define then the 2 propagators $G_{1+}$ and $G_{2+}$ as,
\begin{eqnarray}
G_{1+}(k,\eta)&=&G_{1a}(k,\eta)u_{a}=G_{11}(k,\eta)+G_{12}(k,\eta)\\
G_{2+}(k,\eta)&=&G_{2a}(k,\eta)u_{a}=G_{21}(k,\eta)+G_{22}(k,\eta),
\end{eqnarray}
for which we have,
\begin{eqnarray}
G_{a+}^{\oneloop}(k,\eta)&=&3\,e^{3\eta}\,\int\dd^{3}\vq\,F^{(3)}_{a}(\vq,-\vq,\vk)P_{0}(q)\\
G_{a+}^{\twoloops}(k,\eta)&=&15\,e^{5\eta}\,\int\dd^{3}\vq_{1}
\dd^{3}\vq_{2}\,F^{(5)}_{a}(\vq_{1},-\vq_{1},\vq_{2},-\vq_{2},\vk)P_{0}(q_{1})P_{0}(q_{2})
\end{eqnarray}
with $q_{i}=\vert\vq_{i}\vert$ and where the function $F_{a}^{(m)}$ are the symmetrized PT kernels of the $m$-th order. One can generalize
the above expressions to the one at a general $n$-loop order\footnote{This can be further generalized to the $n$-loop correction to the $p$-point propagator:
$$\Gamma_{\nloops}^{(p)}(\vk_{1},\dots,\vk_{p};\eta)=s_{n}^{(p)}e^{(2n+1)\eta}\int
\dd^{3}\vq_{1}\dots \dd^{3}\vq_{n}\,
F^{(2n+p)}_{a}(\vq_{1},-\vq_{1},\dots,\vq_{n},-\vq_{n},\vk_{1},\dots,\vk_{p})P_{0}(q_{1})\dots P_{0}(q_{n})
$$ with $s_{n}^{(p)}={(2n+p)!(2n-1)!!}/{p!^{2}}$},
\begin{equation}
G_{a+}^{\nloops}=s_{n}^{(1)}\,e^{(2n+1)\eta}\int
\dd^{3}\vq_{1}\dots \dd^{3}\vq_{n}\,
F^{(2n+1)}_{a}(\vq_{1},-\vq_{1},\dots,\vq_{n},-\vq_{n},\vk)P_{0}(q_{1})\dots P_{0}(q_{n})
\end{equation}
 with a symmetry factor given by
 \begin{equation}
s_{n}^{(1)}=\frac{2n+1}{n!}\Pi_{j=1}^{n}\left(\begin{array}{c}
2j\\
2
\end{array}\right)=(2n+1)!!.
\end{equation}
Note that the explicit expressions for the standard PT kernels can be obtained from the recursion relations (see also \cite{2002PhR...367....1B}),
\begin{eqnarray}
F_{a}^{(1)}(\vk_{1})&=&\{1,1\},\\
F_{a}^{(n)}(\vk_{1},\dots,\vk_{n})&=&
\frac{\sigma_{ab}^{(n)}}{n!}\sum_{m=1}^{n-1}\gamma_{abc}(\vq_{1},\vq_{2})F_{c}^{(m)}(\vk_{1},\dots,\vk_{m})F_{d}^{(n-m)}(\vk_{n-m+1},\dots,\vk_{n})
+\hbox{sym.}
\end{eqnarray}
with $\vq_{1}\equiv \vk_{1}+\dots+\vk_{m}$ and $\vq_{2}\equiv \vk_{m+1}+\dots+\vk_{n}$. Here the matrix $\sigma_{ab}^{(n)}$
is given by,
\begin{equation}
\sigma_{ab}^{(n)}=\frac{1}{(2n+3)(n-1)}\left(
\begin{array}{cc}
2n+1& 2\\
3 & 2n
\end{array}
\right),
\label{sigmaab}
\end{equation}
where all the other terms obtained by permutations of the wave modes are included (there are $n!$ of those.).

The resulting expressions depend on the actual linear power spectra. In this sense the result is a priori model dependent.  
Note that in practice we also assume that the structure of the coupling kernels  are that of an Einstein de Sitter universe. In general the coefficients appearing in the expression of Eq. (\ref{sigmaab}) are time dependent. In practice they appear to change very weakly.
As mentioned before, that does not restrict much the scope of these calculations. 

%{\bf To be added: general expression of the p-loop order with symmetry factors, see notes of Atsushi}

The propagators are of further interest because of their high-$k$ behavior that can be derived explicitly taking into account large-scale modes in a full nonlinear manner. Those properties were first derived in 
 \cite{2006PhRvD..73f3520C,2006PhRvD..73f3519C}  from diagram resummations. They were further engrained in a more general framework in \cite{2012PhRvD..85f3509B} with the introduction of the 
 so-called eikonal approximation we will frequently refer to in the following.
From these calculations we expect to have in the high-$k$ limit,
\begin{equation}
G_{ab}(k,\eta)\sim g_{ab}(\eta)\exp\left(-\frac{k^2\sigmad^{2}}{2}(e^{\eta}-1)^2\right),\label{highkGab}
\end{equation}
where $\sigmad$ is the r.m.s. of the displacement field~\footnote{In \cite{2010PhRvD..82h3507B} it is shown how this expression is to be changed in case of primordial non-Gaussian fluctuations.},
\begin{equation}
\sigmad^{2}=\int\frac{\dd^3\vk}{3k^2}P_{0}(k).\label{sigmaddef}
\end{equation}

This expression however does not reproduce the one-loop expression of the propagators. Let us denote $\delta G_{ab}^{\oneloop}$ the expression of the one-loop correction to $G_{ab}$ and $\delta G_{ab}^{\twoloops}$ its expression at two-loop order. 
As shown in \cite{2006PhRvD..73f3519C}, 
it is possible to build an interpolating expression for $G_{ab}$ that reproduces both the asymptotic property
(\ref{highkGab}) and the one-loop order result and which is valid at any time and for any wavenumber $k$. The construction of the interpolation function is however not systematic and relies on specific properties, $k$ and time dependences, of the one-loop expressions.  An alternative form for constructing a systematic interpolation functions of multi-point propagators has been proposed in \cite{2011arXiv1112.3895B}. 
It makes then possible to incorporate corrective terms of arbitrary order into the expression of the propagators. 
We will make use of these results in the following.

%%%%%%%%%%%%%%%%%%%%%%%%%%%%%%%%%%%%%%%%%%%%%%%%%%%%%%%%%%%%%%%%%%%
\subsection{The one-loop order results}

The expressions of the propagator have been  computed in \cite{2006PhRvD..73f3520C} at one-loop order. They appear to be a 
sum of four terms each of which can be factorized 
in time and $k$ dependence. Here and in the following we are interested in the most growing part of the propagators only and for
initial growing modes.
In this regime, we then have,
\begin{eqnarray}
G_{1+}^{\oneloop}(k,\eta)&=&4\pi\,e^{3\eta}\,\int q^2\dd q\,P_{0}(q)\,f(k,q)\label{G11loop}\\
G_{2+}^{\oneloop}(k,\eta)&=&4\pi\,e^{3\eta}\,\int q^2\dd q\,P_{0}(q)\,g(k,q)\label{G21loop}
\end{eqnarray}
with
\begin{eqnarray}
f(k,q)&=&\frac{3 \left(2 k^2+7 q^2\right) \left(k^2-q^2\right)^3 \log \left[\frac{(k-q)^2}{(k+q)^2}\right]+4 \left(6
   k^7 q-79 k^5 q^3+50 k^3 q^5-21 k q^7\right)}{2016 k^3 q^5}\\
g(k,q)&=&\frac{3 \left(2 k^2+q^2\right) \left(k^2-q^2\right)^3 \log \left[\frac{(k-q)^2}{(k+q)^2}\right]+4 \left(6
   k^7 q-41 k^5 q^3+2 k^3 q^5-3 k q^7\right)}{672 k^3 q^5}
\end{eqnarray}
The functions $f(k,q)$ and $g(k,q)$ are dimensionless functions (e.g. they depend only of $k/q$) that are entirely 
determined by the vertices and by the linear propagator. More precisely we have,
\begin{equation}
f(k,q)=\frac{1}{4\pi}\int\dd^2\hq\ F_{3}(\vk,\vq,-\vq),\ \ g(k,q)=\frac{1}{4\pi}\int\dd^2\hq\ G_{3}(\vk,\vq,-\vq),
\end{equation}
where $\hq$ is the relative angle between $\vq$ and $\vk$.

A few remarkable properties are worth mentioning. In particular the low $q$ behaviors of $f(k,q)$ and $g(k,q)$ are directly related to the 
exponential cut-off of large-$k$. We then have
\begin{eqnarray}
f(k,q)\sim g(k,q)\sim-\frac{1}{6}\frac{k^2}{q^2}\ \ \hbox{when}\ q\ll k.
\end{eqnarray}
and the same property is true for $g(k,q)$. The low $k$ properties of $f(k,q)$ and $g(k,q)$ are less universal. There is however
an important result which is due to the low $k$ behavior of the vertices. It implies that
$f(k,q)\sim {k^2}/{q^2}$ and $g(k,q)\sim {k^2}/{q^2}$ when  $k\ll q$. These latter properties ensures a rapid convergence of the integrals in Eqs. (\ref{G11loop},\ref{G21loop}) and are thus important form a practical point of view.
More precisely we have
\begin{equation}
f(k,q)\sim -\frac{61}{630}\frac{k^{2}}{q^{2}}\ \ \hbox{and}\ \ g(k,q)\sim -\frac{3}{10}\frac{k^{2}}{q^{2}}.
\label{eq:fgasymlowk}
\end{equation}
In the next Section we will see how these results are changed for the two-loop results.

%%%%%%%%%%%%%%%%%%%%%%%%%%%%%%%%%%%%%%%%%%%%%%%%%%%%%%%%%%%%%%%%%%%%%%%
\subsection{Multi-loop propagators: constructing regular parts}
\label{sec:regparts} 

Clearly propagators at each order are expected to exhibit specific behavior at large wave number. The leading $k$ behavior is thus expected to behave like,
 \begin{equation}
G_{ab}^{\ploops}(k,\eta)\sim \frac{1}{p!} \frac{(-k)^{2p}}{2^{p}}\sigmad^{2p}(e^{\eta}-1)^{2p}.\label{ploopsd}
\end{equation}
This is precisely what has been obtained in \cite{2006PhRvD..73f3519C} and allows the construction of the form (\ref{highkGab}).
Furthermore, as shown in \cite{2011arXiv1112.3895B} it is possible to write interpolating forms that incorporate both the standard PT results and the expected large-$k$ behaviors of the propagators. At two-loop order this form for the propagator reads 
\begin{equation}
G_{ab}(k,\eta)\approx \left(g_{ab}(\eta)+G_{ab}^{\oneloop}(k,\eta)+\dots
+G_{ab}^{\ploops}(k,\eta)+\ct
\right)\exp\left(-\frac{k^2\sigmad^{2}}{2}(e^{\eta}-1)^2\right)
\label{regG2}
\end{equation}
where $\ct$ is a counter term. It is such that the p-loop order of this form 
reproduces the exact expression of $G_{ab}$ at the same order. For instance a simple calculation leads to
\begin{equation}
\ct=g_{ab}(\eta)\frac{k^2\sigmad^{2}}{2}(e^{\eta}-1)^2+\frac{1}{2}g_{ab}(\eta)\left(\frac{k^2\sigmad^2}{2}\right)^2 (e^{\eta}-1)^2+\frac{1}{2}k^{2}\sigmad^2(e^{\eta}-1)^2
G_{ab}^{\oneloop}(k,\eta).
\end{equation}
for the expression of the counter term up to two-loop order (the first term being its expression at one-loop order). 
 
Here we would like to take advantage of the general properties of the propagators, in particular on its dependence with $\sigmad$, to introduce consistency relations that can be used in practice to test Monte-Carlo
results. The idea is to see $G_{ab}^{\ploops}(k,\eta)$ as a functional of the linear power spectrum $P(k)$ and to express the dependence of 
$G_{ab}^{\ploops}(k,\eta)$ with respect to $\sigmad$.  Let us do this construction order by order. We can define the regular part of 
$G_{ab}^{\oneloop}(k,\eta)$ by
\begin{equation}
G_{ab}^{\oneloop}(k,\eta)=-\frac{k^{2}}{2}\sigmad^{2}(e^{\eta}-1)^{2}g_{ab}(\eta)+\RegG_{ab}^{\oneloop}(k,\eta).
\label{RegOneloops}
\end{equation}
In this expression $\RegG_{ab}^{\oneloop}(k,\eta)$ is the part of $G_{ab}^{\oneloop}(k,\eta)$ which remains finite even when one lets $\sigmad$ go to infinity. The next step is to define the regular part of the two-loop results. The $\sigmad^{4}$ term has already been given in Eq.(\ref{ploopsd}). The $\sigmad^{2}$ term corresponds to the one-loop correction of $\RegG_{ab}^{\oneloop}(k,\eta)$ taken in the high-$k$ limit,
\begin{equation}
G_{ab}^{\twoloops}(k,\eta)=\frac{k^{4}}{8}\sigmad^{4}(e^{\eta}-1)^{4}g_{ab}(\eta)-\frac{k^{2}}{2}\sigmad^{2}(e^{\eta}-1)^{2}\ \RegG_{ab}^{\oneloop}(k,\eta)+\RegG_{ab}^{\twoloops}(k,\eta),
\label{RegTwoloops}
\end{equation}
where again $\RegG_{ab}^{\twoloops}(k,\eta)$ is finite in the formal limit $\sigmad\to\infty$. As we will explicitly use it in the following we go one step further for the three-loop term,
\begin{equation}
\begin{split}
G_{ab}^{\threeloops}(k,\eta)=-\frac{k^{6}}{48}\sigmad^{6}(e^{\eta}-1)^{6}g_{ab}(\eta)
+\frac{k^{4}}{8}\sigmad^{4}(e^{\eta}-1)^{4}\ \RegG_{ab}^{\oneloop}(k,\eta)\\
-\frac{k^{2}}{2}\sigmad^{2}(e^{\eta}-1)^{2}\ \RegG_{ab}^{\twoloops}(k,\eta)+\RegG_{ab}^{\threeloops}(k,\eta).
\label{RegThreeloops}
\end{split}
\end{equation}
In this expression the first term is obtained from the large $k$ behavior of the three-loop expression, the second from the
large-$k$ behavior of the four-point propagator and the third from the six-point propagator computed taken in configurations
where all modes are in the high $k$ limit.
 
The expression (\ref{RegThreeloops}) can be used as a consistency checks for Monte Carlo integrations of three-loop contributions. Indeed one expects the last term $\RegG_{ab}^{\threeloops}(k,\eta)$ to then remain finite. And from a $\sigmad$ expansion point of view the expression (\ref{regG2}) simply reads,
\begin{equation}
G_{ab}(k,\eta)\approx \left(g_{ab}(\eta)+\RegG_{ab}^{\oneloop}(k,\eta)+\RegG_{ab}^{\twoloops}(k,\eta)\right)
\exp\left(-\frac{k^2\sigmad^{2}}{2}(e^{\eta}-1)^2\right).
\label{regprop2}
\end{equation}
This form underlines the importance of the regular part of each contributing term we have defined. 
 
Note that in the previous calculations the expression of $\sigmad$ is computed a priori from linear theory, i.e. Eq. (\ref{sigmaddef}), but its computation can incorporate in principle higher order terms. This is a direct consequence of the eikonal results where it is shown that 
the variance of the displacement field can be computed beyond the linear regime\footnote{That the displacement field could be exponentiated beyond linear order was also noticed in \cite{2011JCAP...06..015A}}. For instance the 1-loop correction to the displacement  field can be explicitly computed and it reads,
\begin{eqnarray}
\sigma_{{\rm d}, \oneloop}^{2}&=&\frac{(4\pi)^{2} }{112896}\int \frac{\dd q_{1}}{q_{1}^{3}}P_{0}(q_{1})\,\frac{\dd q_{2}}{q_{2}^{3}}P_{0}(q_{2})\,
\nonumber\\
&&\times \left[\left(q_1^2-q_2^2\right){}^4 \log
   \left(\frac{\left(q_1+q_2\right){}^2}{\left(q_1-q_2\right){}^2}\right)-4 q_1 q_2 \left(q_1^2+q_2^2\right)
   \left(q_1^4-70 q_2^2 q_1^2+q_2^4\right)\right].
    \label{sigmad1l}
\end{eqnarray}
In such a case the  corrective term can be taken into account in the expression of the regular part. Eq. (\ref{RegOneloops}) is then left unchanged but Eq. (\ref{RegTwoloops}) is modified into 
\begin{eqnarray}
G_{ab}^{\twoloops}(k,\eta)&=&-\frac{k^{2}}{2}\sigma_{{\rm d},\oneloop}^{2}e^{2\eta}(e^{\eta}-1)^{2}g_{ab}(\eta)+\frac{k^{4}}{8}\sigmadlin^{4}(e^{\eta}-1)^{4}g_{ab}(\eta)-\frac{k^{2}}{2}\sigmadlin^{2}(e^{\eta}-1)^{2}\ \RegG_{ab}^{\oneloop}(k,\eta)\nonumber\\
&&+\RegG_{ab}^{\twoloops}(k,\eta),
\end{eqnarray}
in order to incorporate the one-loop correction that appears in the counter term of $\RegG_{ab}^{\oneloop}(k,\eta)$. Then 
Eq. (\ref{regprop2}) is changed into,
\begin{equation}
G_{ab}(k,\eta)\approx \left(g_{ab}(\eta)+\RegG_{ab}^{\oneloop}(k,\eta)+\RegG_{ab}^{\twoloops}(k,\eta)\right)
\exp\left(-\frac{k^2(\sigmadlin^{2}+\sigma_{{\rm d},\oneloop}^{2}e^{2\eta})}{2}(e^{\eta}-1)^2\right),
\label{regprop3}
\end{equation}
when the propagator is written up to two-loop order. Note that the extension to the three-loop terms is a bit more cumbersome. This is due to the fact that  it is then necessary to take into account  fourth order cumulant, the trispectrum, of the displacement field that appear at this order of PT calculations which should also be incorporated in the exponential factor\footnote{an aspect that was somewhat overlooked in \cite{2011JCAP...06..015A}.} as explicitly shown in \cite{2010PhRvD..82h3507B}. 

The importance of the change from (\ref{regprop2}) to (\ref{regprop3}) will be discussed in the next section from the explicit computations of the two-order expression. As we will see the changes it introduces are very benign. In practice we use an exponential factor which is 
computed either from linear theory or from explicit measurements in simulations. It is also to be noted that the concept of regular parts can be extended to  other types of PT diagrams, not necessarily those contributing to the two-point propagators. We leave this analysis for a further study.
%{\bf FB : might be interesting to see how results are changed though.}

 %%%%%%%%%%%%%%%%%%%%%%%%%%%%%%%%%%%%%%%%%%%%%%%%%%%%%%%%%%%%%%%%%%%%%%
\section{Two-loop results}
\label{sec:thediagrams}

\subsection{The contributing diagrams}

\begin{figure}[ht] %  figure placement: here, top, bottom, or page
   \centering
 \includegraphics[width=6cm]{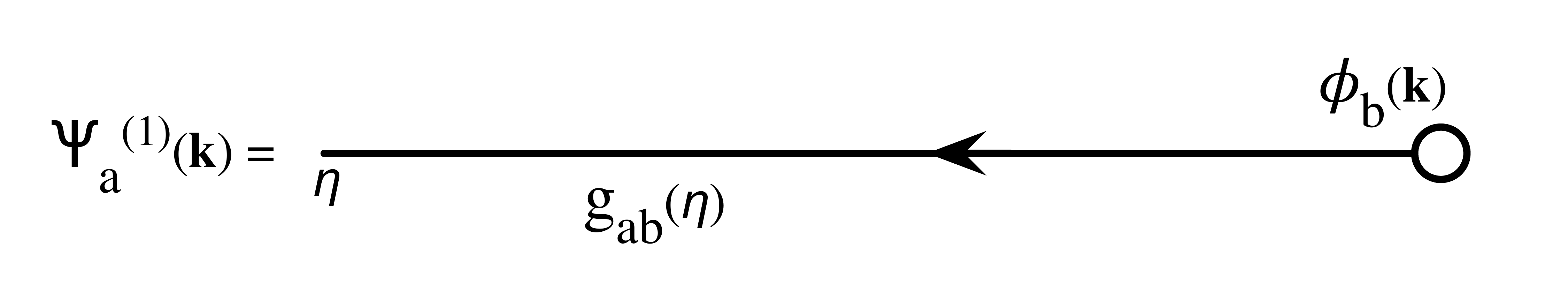} 
   \caption{Diagrammatic representation of the linear propagator. $\phi_{b}$ represents the initial conditions and $g_{ab}$ is the time dependent propagator. This diagram value is given by Eq. (\ref{eomi}) when the second term of the right hand side has been dropped.}
   \label{PsiLinear}
\end{figure} 

\begin{figure}[ht] %  figure placement: here, top, bottom, or page
   \centering
 \includegraphics[width=6cm]{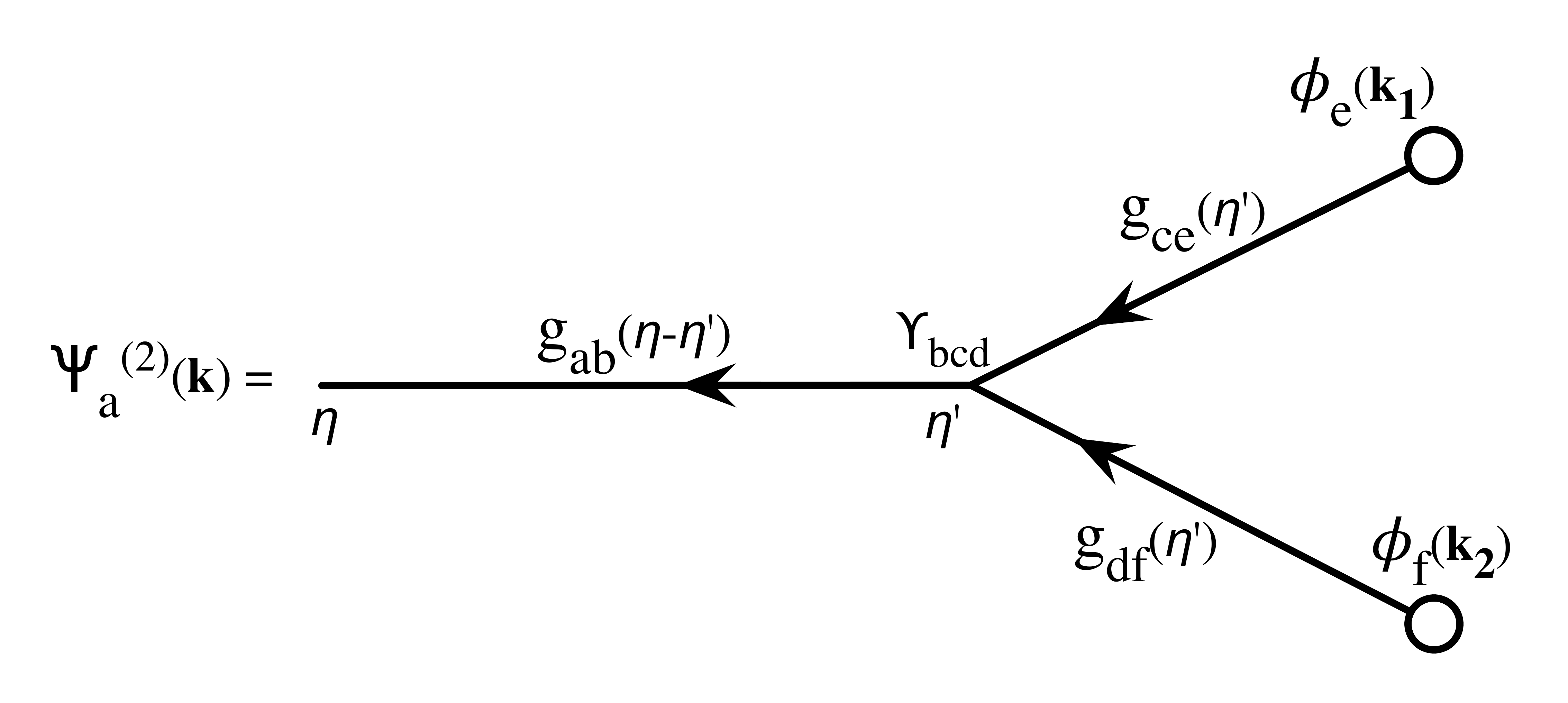} 
   \caption{Diagrammatic representation of the fields at second order. This diagram value is given by Eq. (\ref{eomi}) when one replaces $\Psi_{c}$ and $\Psi_{d}$ in the second term of the right hand by their linear expressions. In the diagram, each time one encounters a vertex, a time integration and a Dirac delta function in the wave modes is implicitly assumed.}
   \label{Psi2Order}
\end{figure} 

\begin{figure}[ht] %  figure placement: here, top, bottom, or page
   \centering
 \includegraphics[width=8cm]{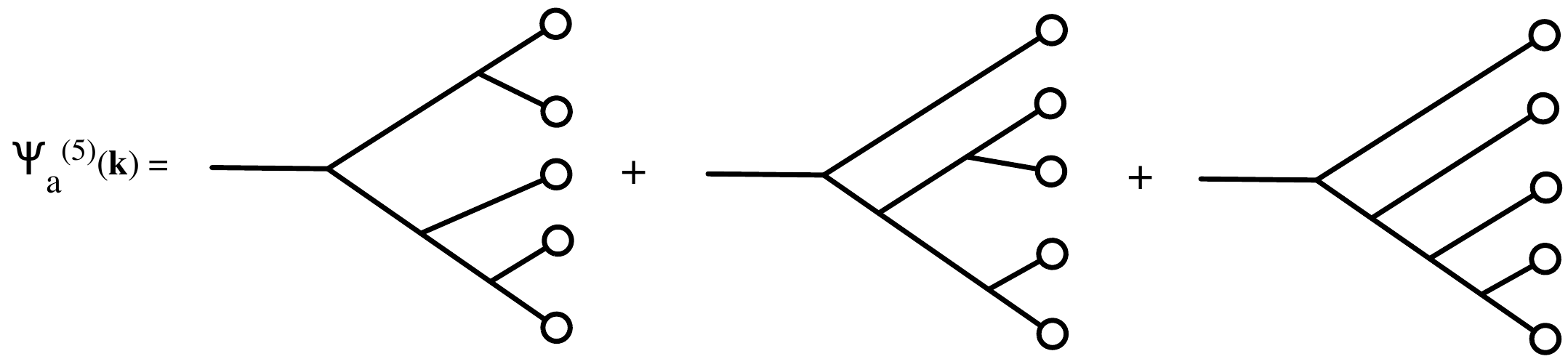} 
   \caption{Diagrammatic representation of the fields at fifth order. Three different diagrams are found to contribute.}
   \label{Psi5Order}
\end{figure}

The computation of $G_{ab}^{\twoloops}$ is still an arduous task that relies on the computation of many contributing terms. 
These terms can be formally obtained from the expression of $\Psi_{a}^{(5)}$ that is the expression of the cosmic fields at fifth order in the initial conditions. Formally the resulting expression $G_{a+}^{\twoloops}$ can be written,
\begin{eqnarray}
G_{1+}^{\twoloops}(k,\eta)&=&(4\pi)^2\ e^{5\eta}\,\int q_{1}^2 \dd q_{1 }\,P_{0}(q_{1}) \int q_{2}^2 \dd q_{2 }\,P_{0}(q_{2})\ \vf(k,q_{1},q_{2})\label{vfDef}\\
G_{2+}^{\twoloops}(k,\eta)&=&(4\pi)^2\ e^{5\eta}\,\int q_{1}^2 \dd q_{1 }\,P_{0}(q_{1}) \int q_{2}^2 \dd q_{2 }\,P_{0}(q_{2})\ \vg(k,q_{1},q_{2})\label{vgDef}.
\end{eqnarray}
where the expressions of the functions $\vf$ and $\vg$ can be derived from $F_{5}$ and $G_{5}$ respectively,
\begin{eqnarray}
\vf(k,q_{1},q_{2})&=&\frac{1}{(4\pi)^2}\int\dd^2\hq_{1}\dd^2\hq_{2}\ F_{5}(\vk,\vq_{1},-\vq_{1},\vq_{2},-\vq_{2}),\label{vfexp}\\
\vg(k,q_{1},q_{2})&=&\frac{1}{(4\pi)^2}\int\dd^2\hq_{1}\dd^2\hq_{2}\ G_{5}(\vk,\vq_{1},-\vq_{1},\vq_{2},-\vq_{2}),\label{vgexp}
\end{eqnarray}
Interestingly these expressions can be represented at a diagrammatic level.

\begin{figure}[ht] %  figure placement: here, top, bottom, or page
   \centering
 \includegraphics[width=9cm]{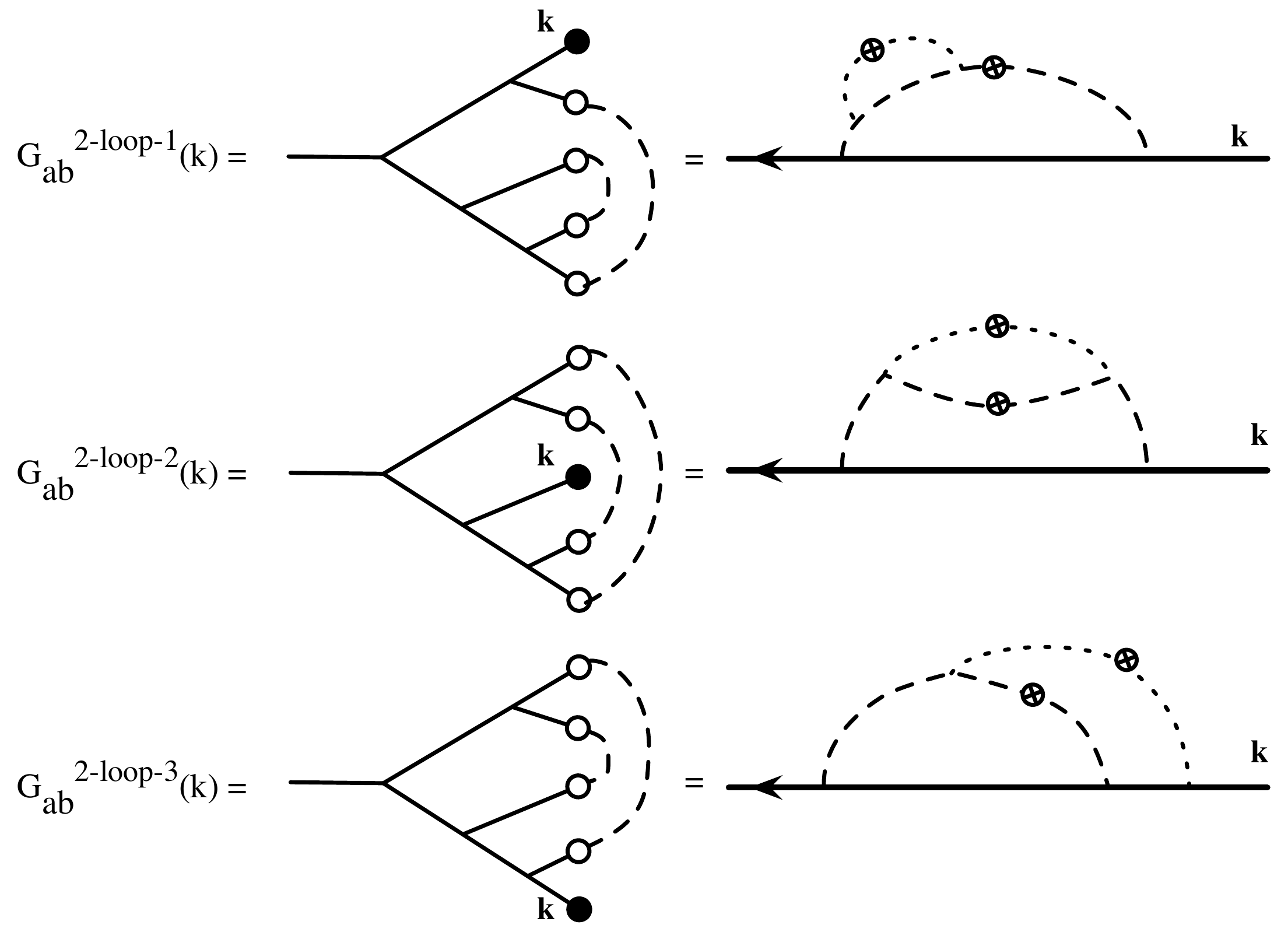} 
 \includegraphics[width=9cm]{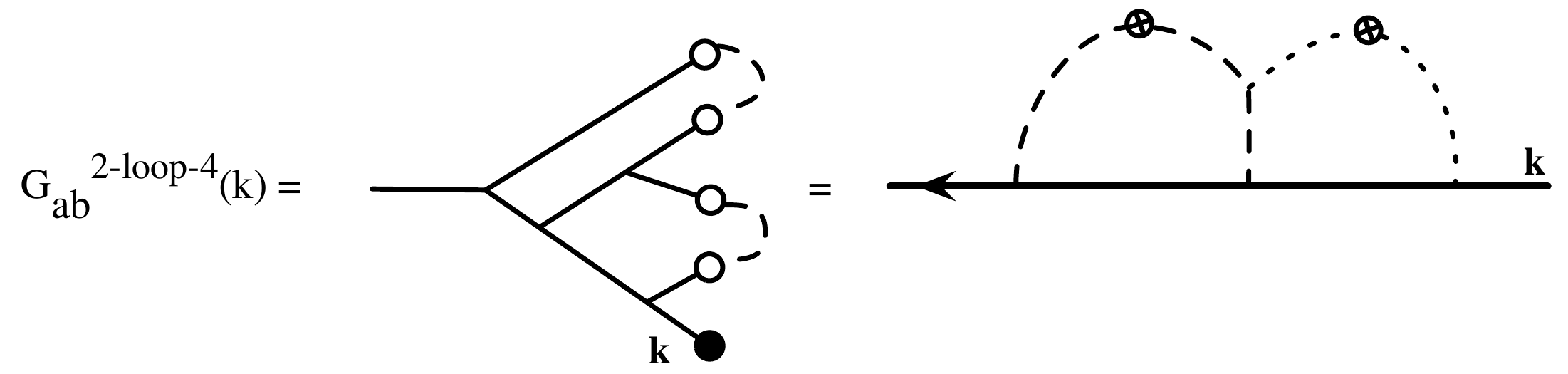} 
 \includegraphics[width=9cm]{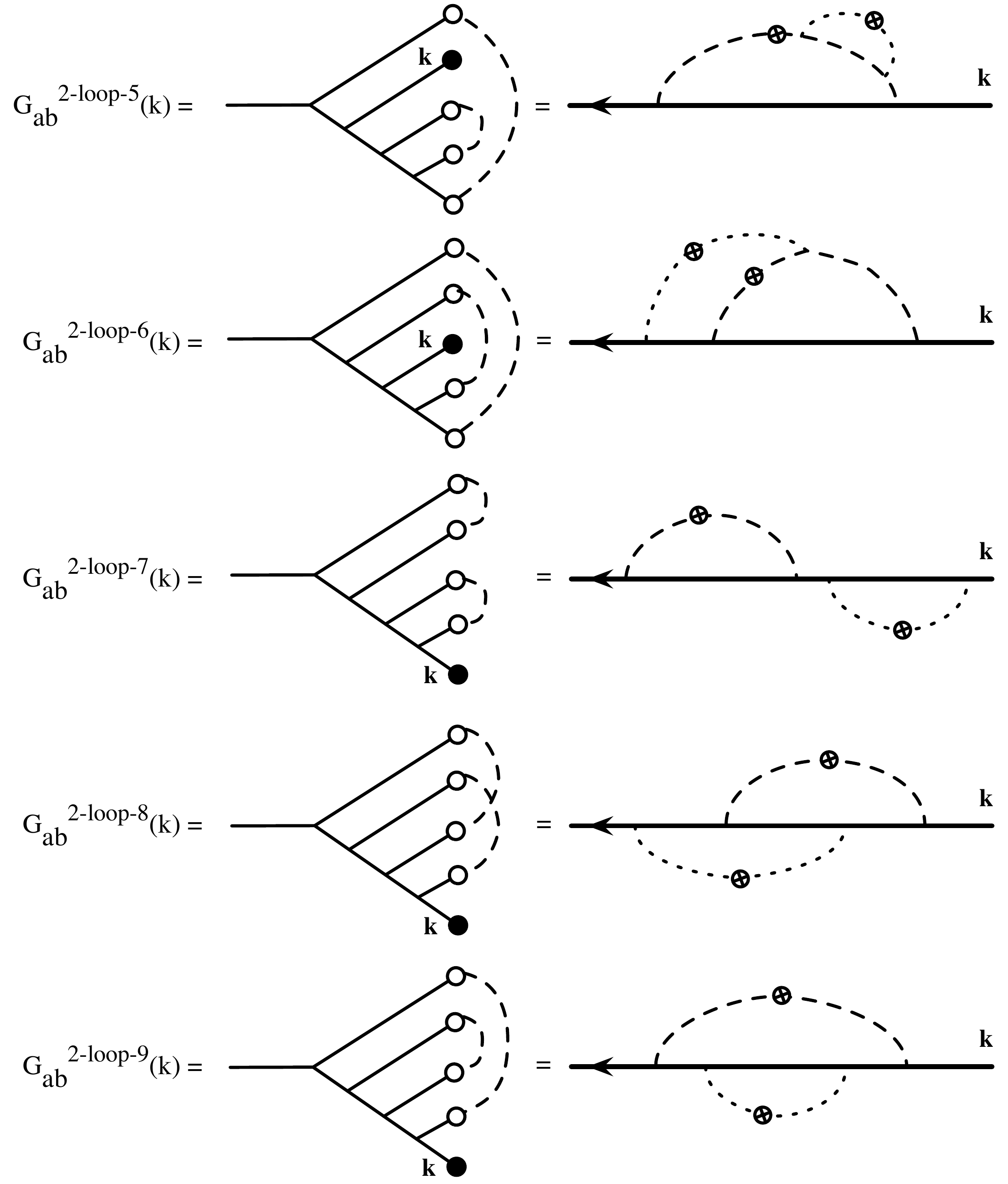} 
   \caption{Diagrams contributing to the two-loop expression of the propagators.}
   \label{diagrams}
\end{figure}

In Fig. \ref{PsiLinear} 
defines the linear propagator. In Fig. \ref{Psi2Order} we show how the coupling term is represented. In this diagram we show the formal expression of the second order fields. And Finally, in Fig. \ref{Psi5Order} we represent the 3 types of diagrams that contribute to the fifth order expression of the fields. It is this one that one should use to construct the two-loop contributing terms. 
They are obtained when 2 of the incoming lines are ``glued'' together: each time it introduces a linear power spectrum factor represented by a crossed circle. We are then led to the contributing diagrams depicted on Fig. \ref{diagrams}.
At this stage diagrams represent a mere transcription of the expansion of the motion equations.
A detailed description of the procedure to draw the diagrams and compute their values can be found
in \cite{2006PhRvD..73f3519C}.

There appear to be 9 different diagrams contributing to
$\vf$ and $\vg$. Note that each of these diagrams comes with a symmetry factor. The computation of these symmetry factors 
can be done by counting the number of possibilities one has to get the same final expression out of a given contributing diagram to $\Psi^{(5)}$. Finally the symmetry factors of the diagrams of Fig.  \ref{diagrams} are respectively, $2,1,2,4,1,1,1,1,1$ for diagrams 1 to 9. Note that the diagrams contributing to the one-loop correction of the displacement field, Eq. (\ref{sigmad1l}), are the diagrams 1, 2 and 5 (with their previous symmetry factors) when computed in the eikonal limit.

A priori the 2 angular integrations contained in Eqs. (\ref{vfexp}-\ref{vgexp}) can be done irrespectivevely of the details of the shape of the power
spectrum and can therefore be done in a model independent way. 
The whole computation is however cumbersome and leads to a very large number of terms, the complexity of which depends on the diagrams. Some diagrams correspond to ``simple'' diagram (the diagrams 1, 5, 7 and 9) the expression of which can be expressed in terms of the function $f$ or $g$ only. This is not the case however for the others. Such angular integrations involve various sets of functions. One is built out of
\begin{eqnarray}
\int\dd^{2}\hq\,\frac{1}{\vert \vk+\vq\vert^2}= \frac{\pi}{k q}   \log\frac{(k+q)^2}{(k-q)^2},
\label{Ha00}
\end{eqnarray}
and another of
\begin{equation}
\int\dd^{2}\hq\,\frac{1}{\vert \vk_{1}+\vq\vert^2\,\vert \vk_{2}-\vq\vert^2}=
\frac{\pi \log\left(-\mW_{+}/\mW_{-}\right)}{k_3\, q\,
\sqrt{\left(k_2^2-q^2\right)
   k_1^2+q^2 \left(-k_2^2+k_3^2+q^2\right)}}
   \label{Hd00}
\end{equation}
with
\begin{eqnarray}
\mW_{\pm}\equiv \pm 2(k_1^2-q^2)(k_2^2-q^2) \pm 4 k_3^2 q^2+ 4 k_3 q \sqrt{k_1^2 k_2^2 + q^4 + q^2(k_3^2-k_1^2-k2^2)}. 
\end{eqnarray}
These two functions are those that appear in the expression of the one-loop correction to $\Gamma^{(2)}$. The calculation 
of $\vf$ and $\vg$ involve however further integrals of the form,
\begin{eqnarray}
\int\dd^{2}\hq_{1}\dd^{2}\hq_{2}\,\frac{1}{\vert \vq_{1}+\vk\vert^2\vert \vq_{2}+\vk\vert^2\vert \vq_{1}+\vq_{2}+\vk\vert^2}
\end{eqnarray}
(for diagram 5),
\begin{eqnarray}
\int\dd^{2}\hq_{1}\dd^{2}\hq_{2}\,\frac{1}{\vert \vq_{1}+\vq_{2}\vert^4\vert \vq_{1}+\vq_{2}+\vk\vert^2}
\end{eqnarray}
 (for diagram 6),
\begin{eqnarray}
\int\dd^{2}\hq_{1}\dd^{2}\hq_{2}\,\frac{1}{\vert \vq_{1}+\vq_{2}\vert^2\vert \vq_{1}+\vk\vert^2\vert \vq_{1}+\vq_{2}+\vk\vert^2},
\end{eqnarray}
(for diagrams 7 and 8) or 
\begin{eqnarray}
\int\dd^{2}\hq_{1}\dd^{2}\hq_{2}\,\frac{1}{\vert \vq_{1}+\vk\vert^2\vert \vq_{2}-\vk\vert^2\vert \vq_{1}+\vq_{2}\vert^2}
\end{eqnarray}
(for diagram 9). The angular integrations in those expressions can be done only partially using the relations (\ref{Ha00}, \ref{Hd00}).
We are left with expression that involve one single angle (that could indifferently 
be the one between $\vq_{1}$ and $\vq_{2}$ or $\vk$ or the angle between $\vq_{2}$ and $\vk$ depending on the choice of order of 
angle integrations).
The resulting expressions are yet far too complicated to be reproduced here. In the following we rather give some general asymptotic properties and propose some fitting functions.

\subsection{The asymptotic behaviors}
\label{sec:props}

\begin{figure}[ht] %  figure placement: here, top, bottom, or page
   \centering
   \includegraphics[width=12cm]{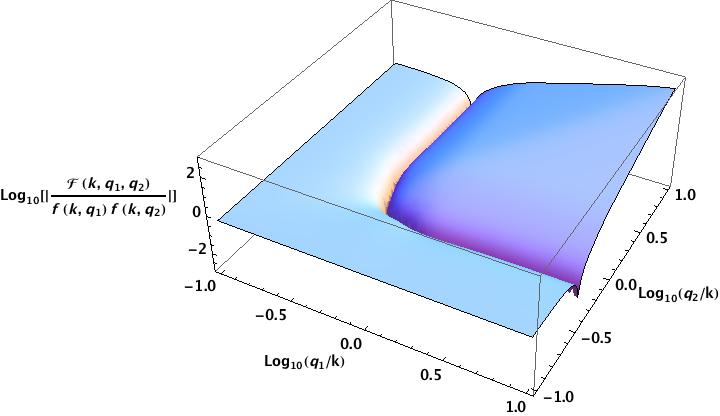} %\hspace{.5cm}
      \includegraphics[width=12cm]{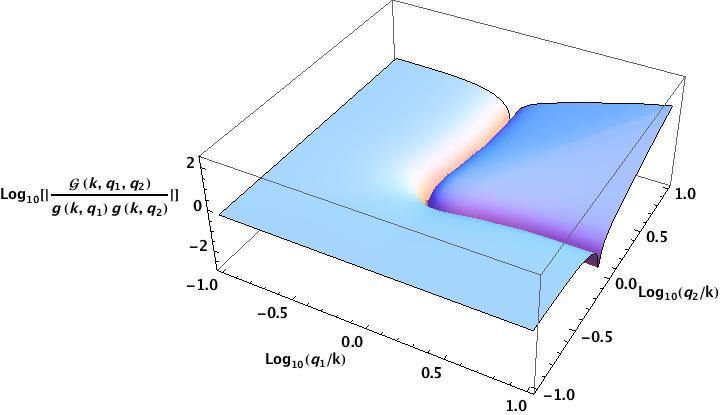} 
   \caption{The functions $\vf(k,q_{1},q_{2})$ and $\vg(k,q_{1},q_{2})$ in units respectively of $f(k,q_{1})f(k,q_{2})$ and $g(k,q_{1})g(k,q_{2})$.
   This illustrates the validity of the asymptotic expansions of Eq. (\ref{vfvgexpansions}).}
   \label{vfvgshapes}
\end{figure}

The high $k$ asymptotic properties of $\vf$ and $\vg$ can be inferred from the fact that $G_{ab}^{\twoloops}$ should be consistent with the second order term of expression (\ref{highkGab}) in a perturbative  expansion. That implies in particular that 
\begin{equation}
\vf(k,q_{1},q_{2})\to \frac{1}{2}\frac{k^2}{6 q_{1}^2}\frac{k^2}{6 q_{2}^2}.
\end{equation}
This asymptotic property can actually be extended in a more subtle way. Let us assume that both $k$ and $q_{2}$ are finite and that $q_{1}$ is much smaller than both $q_{2}$ and $k$. In this case $\vf(q_{1},q_{2},k)$ is nothing but the one-loop correction of $\Gamma^{(3)}$ taken in a particular configuration. In this limit though the one-loop correction is independent  on the configuration  and is given by $-k^2/(6q_{1}^2)$ times
the tree order result. As a consequences one gets a relation between $\vf$ and $f$,
\begin{equation}
\vf(k,q_{1},q_{2})\to -\frac{k^2}{12q_{1}^2}f(k,q_{2})\ \ \hbox{when}\ \ q_{1}\ll k \hbox{ and } q_{1}\ll q_{2}.
\end{equation}
A similar expression naturally holds when the roles of $q_{1}$ and $q_{2}$ are inverted. These two results can be recapped in the following form,
\begin{equation}
\vf(k,q_{1},q_{2})\to \frac{1}{2}f(k,q_{1})f(k,q_{2})\ \ \hbox{when}\ \ q_{1}\ll k \hbox{ or } q_{2}\ll k.\label{vfvgexpansions}
\end{equation}
Note that The validity of this approximation is controlled by the magnitude of $f(k,q_{1})+f(k,q_{2})$. Moreover when either $q_{1}$ or $q_{2}$ is large one expects to obtain either a $1/q_{1}^2$ or $1/q_{2}^2$ behavior respectively. This comes from the IR properties of the kernels. All these properties can be encapsulated in the following forms,
\begin{eqnarray}
\vf(k,q_{1},q_{2})&=&\frac{1}{2}f(k,q_{1})f(k,q_{2})-\frac{k^2}{q_{1}^2+q_{2}^2}\alpha_{f}(q_{1}/k,q_{2}/k),
\label{alphafdef}\\
\vg(k,q_{1},q_{2})&=&\frac{1}{2}g(k,q_{1})g(k,q_{2})-\frac{k^2}{q_{1}^2+q_{2}^2}\alpha_{g}(
q_{1}/k,q_{2}/k),\label{alphagdef}
\end{eqnarray}
where $\alpha_{f}$ and $\alpha_{g}$ are both finite. The first term dominates when either $q_{1}$ or $q_{2}$ is small, the second when both are large. Note that the second term is always negative. This is illustrated on Fig. \ref{vfvgshapes}.

\begin{figure}[ht] %  figure placement: here, top, bottom, or page
   \centering
   \includegraphics[width=12cm]{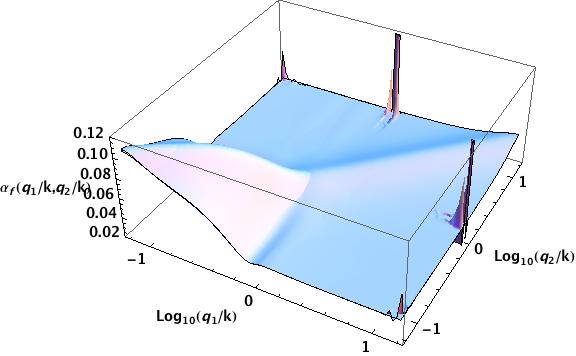}
      \includegraphics[width=12cm]{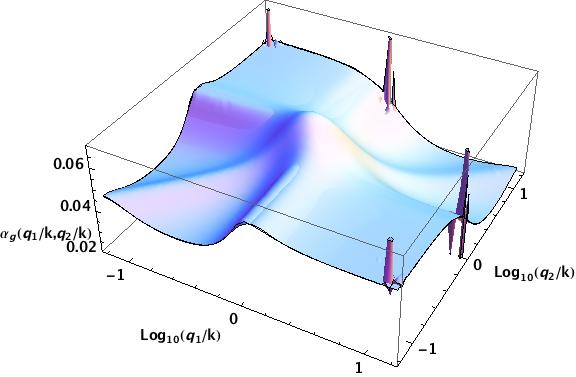} 
   \caption{The functions $\alpha_{f}(k,q_{1},q_{2})$ and $\alpha_{g}(k,q_{1},q_{2})$. Each plot shows the superposition of the result of the numerical integration and the fitted functions. The results of the numerical integrations show numerical instabilities that the fitted functions do not have. The relative error between the the fitted functions and the results of the numerical integrations is less than 1\% when the latter is reliable.}
   \label{alphafgshapes}
\end{figure}

The functions $\alpha_{f}$ and $\alpha_{g}$ defined in Eqs. (\ref{alphafdef}, \ref{alphagdef}) are shown on Fig. \ref{alphafgshapes}. 
They are clearly finite in all possible $q_{1}$ and $q_{2}$ ranges. It is actually to compute their asymptotic expressions when 
$k$ is small. They read,
\begin{equation}
\alpha_{f}(q_{1},q_{2})\ \to\ \beta_{f}(q_{1},q_{2})\ \ \hbox{and}\ \ \alpha_{g}(q_{1},q_{2})\ \to\ \beta_{g}(q_{1},q_{2})
\end{equation}
with
\begin{eqnarray}
\beta_{f}(q_{1},q_{2})&=&
\frac{\left(q_1^2+q_2^2\right)}{27518400 q_1^7 q_2^7}
 \Big\{4 q_1 q_2 \left(5760 q_1^{10}+19365 q_2^2 q_1^8-114653 q_2^4 q_1^6-114653 q_2^6 q_1^4+19365 q_2^8
   q_1^2+5760 q_2^{10}\right)\nonumber\\
&&   +15 \left(384 q_1^4+2699 q_2^2 q_1^2+384 q_2^4\right) \left(q_1^2-q_2^2\right){}^4 \log
   \left[\frac{\left(q_1-q_2\right){}^2}{\left(q_1+q_2\right){}^2}\right]\Big\}
\end{eqnarray}
and
\begin{eqnarray}
\beta_{g}(q_{1},q_{2})&=&
\frac{\left(q_1^2+q_2^2\right)}{302702400 q_1^7 q_2^7} \Big\{4 q_1 q_2 \left(17280 q_1^{10}+120495 q_2^2 q_1^8-572759 q_2^4 q_1^6-572759 q_2^6 q_1^4+120495 q_2^8
   q_1^2+17280 q_2^{10}\right)
   \nonumber\\
&&   +15 \left(1152 q_1^4+12257 q_2^2 q_1^2+1152 q_2^4\right) \left(q_1^2-q_2^2\right){}^4 \log
   \left[\frac{\left(q_1-q_2\right){}^2}{\left(q_1+q_2\right){}^2}\right]\Big\}.
\end{eqnarray}
These forms are reached whenever both $q_{1}/k$ and $q_{2}/k$ are large.

In practice however the previous formulation can lead to numerical instabilities when the ratio $q_{1}/q_{2}$ is large. Alternative forms that
make use of the expansion of the logarithm when its variable is close to unity can be employed. They are of the form,
\begin{eqnarray}
\beta_{f}(q_{1},q_{2})&=&\frac{\left(q_1^2+q_2^2\right)}{27518400 q_1^7 q_2^7 \left(q_1+q_2\right){}^4}
 \left(15 \left(q_1-q_2\right){}^4 \left(q_1+q_2\right){}^8 \left(384 q_1^4+2699 q_2^2 q_1^2+384 q_2^4\right)
   B_{\nu}(5,0)
  \right.\nonumber\\
&&   \left.\hspace{-2cm} -512 q_1^5 q_2^5 \left(2304 q_1^6-6144 q_2 q_1^5+18103 q_2^2 q_1^4-50908 q_2^3 q_1^3+18103
   q_2^4 q_1^2-6144 q_2^5 q_1+2304 q_2^6\right)\right)
\end{eqnarray}
and
\begin{eqnarray}
\beta_{g}(q_{1},q_{2})&=&
\frac{\left(q_1^2+q_2^2\right)}{302702400 q_1^7 q_2^7 \left(q_1+q_2\right){}^4} 
\left(15 \left(q_1-q_2\right){}^4 \left(q_1+q_2\right){}^8 \left(1152 q_1^4+12257 q_2^2 q_1^2+1152 q_2^4\right)
   B_{\nu}(5,0)\right.\nonumber\\
&&   \left.  \hspace{-2cm}-512 q_1^5 q_2^5 \left(6912 q_1^6-18432 q_2 q_1^5+75109 q_2^2 q_1^4-235924 q_2^3
   q_1^3+75109 q_2^4 q_1^2-18432 q_2^5 q_1+6912 q_2^6\right)\right)
\end{eqnarray}
where $B_{\nu}(5,0)$ is the incomplete Beta function of index $\nu=\frac{4 q_1 q_2}{\left(q_1+q_2\right){}^2}$.

\subsection{Fitted forms}

As it is nearly impossible to reproduce the explicit result for the functions $\vf$ and $\vg$, we rather propose some analytical fits for 
$\alpha_{f}$ and $\alpha_{g}$ that are given below.

They can be fitted with the form (where $a=f$ or $a=g$),
\begin{equation}
\alpha_{a}(q_{1},q_{2})=\left(\beta_{a}(q_{1},q_{2})/b_{a}-c_{a}(q_{s})-d_{a}(q_{r})\right)\delta_{a}(q_{s})\\
\end{equation}
where $b_{a}$ is the value of $\beta_{a}$ for $q_{1}=q_{2}$,
\begin{equation}
b_{f}=\frac{22382}{429975}\ \ \hbox{and}\ \ b_{g}=\frac{9886}{429975},
\end{equation}
the variable $q_{s}$ and $q_{r}$ are defined by
\begin{equation}
q_{r}=\sqrt{q_{1}^2+q_{2}^2}\ \ \hbox{and}\ \ q_{s}=1/\sqrt{q_{1}^{-2}+q_{2}^{-2}}.
\end{equation}
The functions $c_{a}(q_{s})$, $d_{a}(q_{r})$ and $\delta_{a}$ are fitted functions.
They are given by,
\begin{eqnarray}
c_{f}(q_{s})&=&\frac{0.401228}{35.0987 q_s^4+4.13381 q_s^2+1}-\frac{1.46195 q_s^3}{77.7967 q_s^6+1}\\
d_{f}(q_{r})&=&\frac{0.178036 q_r^2}{4.11634 q_r^{10}+1}-\frac{0.425873}{7.03063 q_r^4+2.45787 q_r^2+1}\\
\delta_{f}(q_{s})&=&
-\frac{0.0419954 \left(76.7666 q_s^4+4.33008 q_s^2\right)}{76.7666 q_s^4+1}+\frac{0.0121732}{1010.66 q_s^4+28.3738 q_s^2+1}+0.0941118\\
c_{g}(q_{s})&=&\frac{0.478032 q_s^{3.}}{28.4904 q_s^8+1}-\frac{0.357391}{42.1407 q_s^4+1.36756 q_s^2+1}\\
d_{g}(q_{r})&=&\frac{0.348953}{2.23826 q_r^5+1}-\frac{0.268523 q_r^2}{0.471174 q_r^5+1}\\
\delta_{g}(q_{s})&=&
\frac{0.0351332 \left(9.76583 q_s^4+3.72325 q_s^2\right)}{9.76583 q_s^4+1}+\frac{0.0580042}{0.0112425 q_s^4+2.06647 q_s^2+1}-0.0120398
\end{eqnarray}

\subsection{The resulting propagators}
\label{subsec:resprop}

The resulting propagators are computed with the help of the Eqs. (\ref{vfDef}, \ref{vgDef}). 
The regular part of $G^{\twoloops}$ is then given by
\begin{eqnarray}
\RegG_{a+}^{\twoloops}(k,\eta)=\frac{1}{2}\left[\RegG^{\oneloop}(k,\eta)\right]^{2}
-(4\pi)^{2}\int{\dd q_{1}}\,{\dd q_{2}}\,P(q_{1})\,P(q_{2})\,\frac{q_{1}^{2}\,q_{2}^{2}\,k^{2}}{q_{1}^{2}+q_{2}^{2}}\ 
\alpha_{a}(q_{1}/k,q_{2}/k).
\end{eqnarray}
 It is to be noted that the integrals over 
$q_{1}$ and $q_{2}$ safely converge for $\Lambda$-CDM type models. For a power spectrum that reaches a power law behavior of index $n$ at large $k$ one can easily check that the convergence is obtained for $n<-2$ (the condition is $n<-1$ for the one-loop expression).
These convergence properties will be discussed in more detail in the next Section.

Finally we present the resulting shape of the propagators.
At one- and two-loop level the regularized approach is un-ambiguous. It leads to,
\begin{equation}
G_{a+}^{\reg-\oneloop}(k,\eta)=\left[e^{\eta}+\RegG_{a+}^{\oneloop}(k,\eta)\right]
\exp\left(-\frac{1}{2}k^{2}\sigmad^{2}\,e^{2\eta}\right).\label{GaplusReg1}
\end{equation}
and
\begin{equation}
G_{a+}^{\reg-\twoloops}(k,\eta)=\left[e^{\eta}+\RegG_{a+}^{\oneloop}(k,\eta)+\RegG_{a+}^{\twoloops}(k,\eta)\right]
\exp\left(-\frac{1}{2}k^{2}\sigmad^{2}\,e^{2\eta}\right).\label{GaplusReg2}
\end{equation}

\begin{figure}[t] %  figure placement: here, top, bottom, or page
   \centering
   \includegraphics[width=7cm]{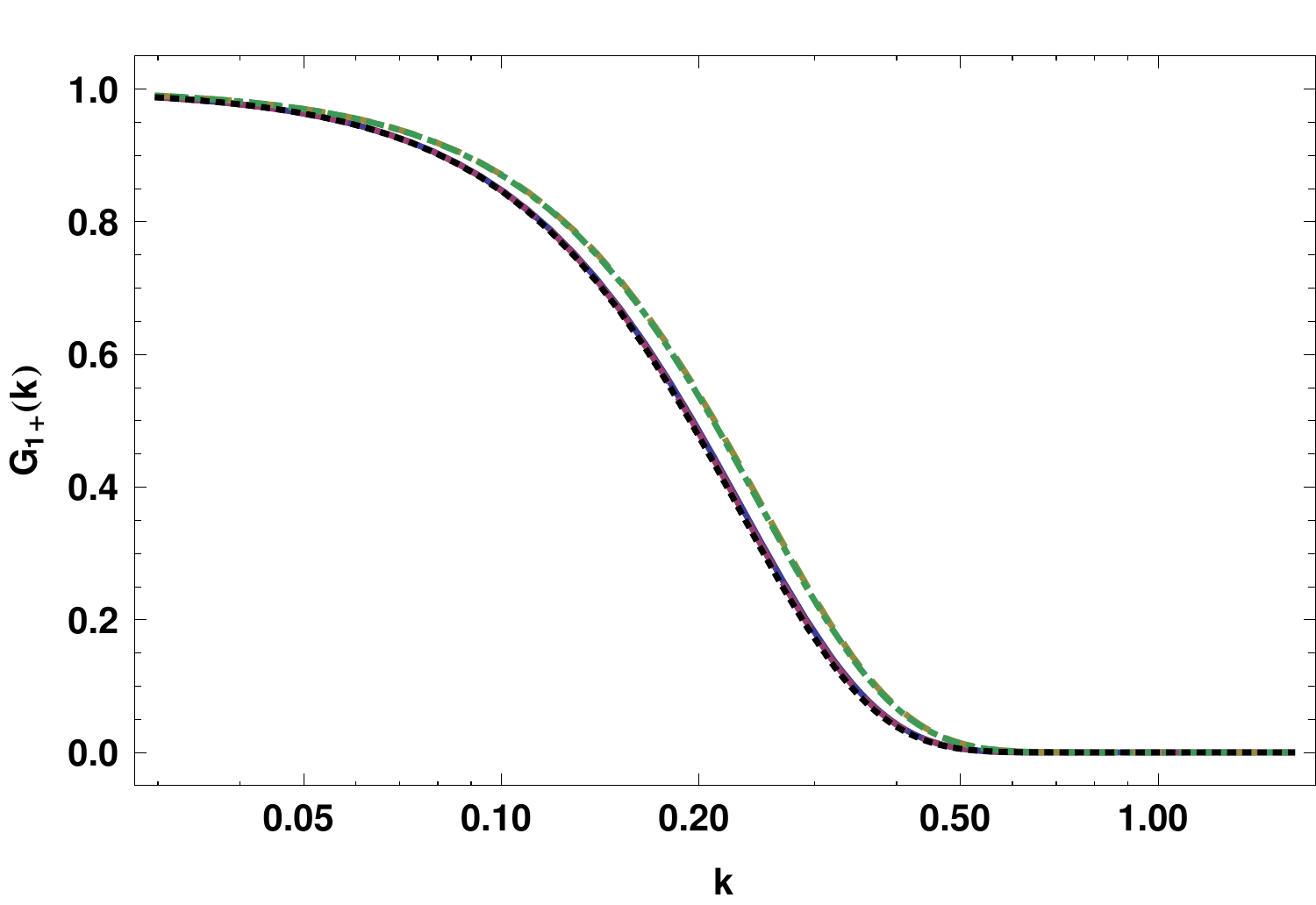} \hspace{.5cm}  
   \includegraphics[width=7cm]{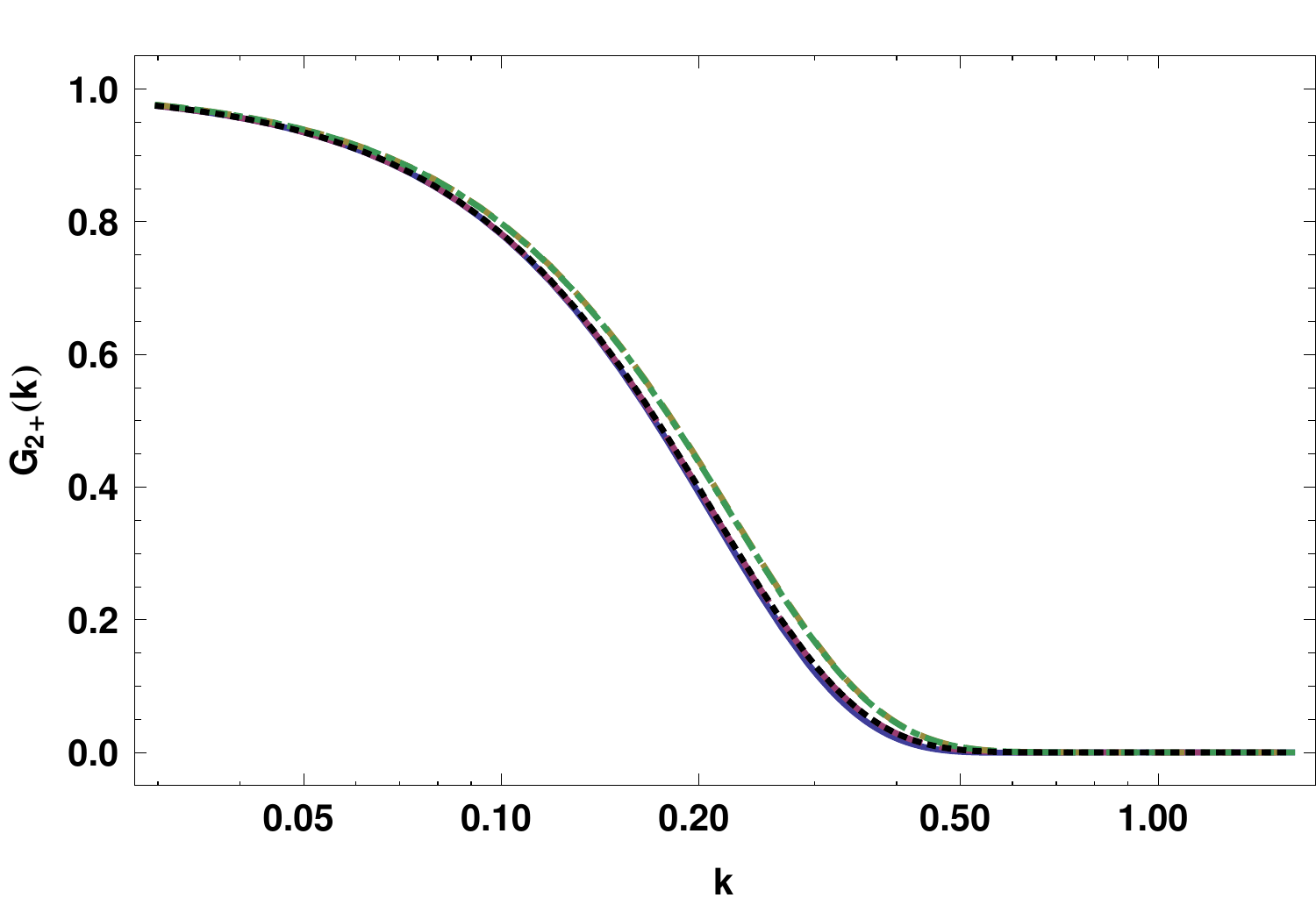} \\
      \includegraphics[width=7cm]{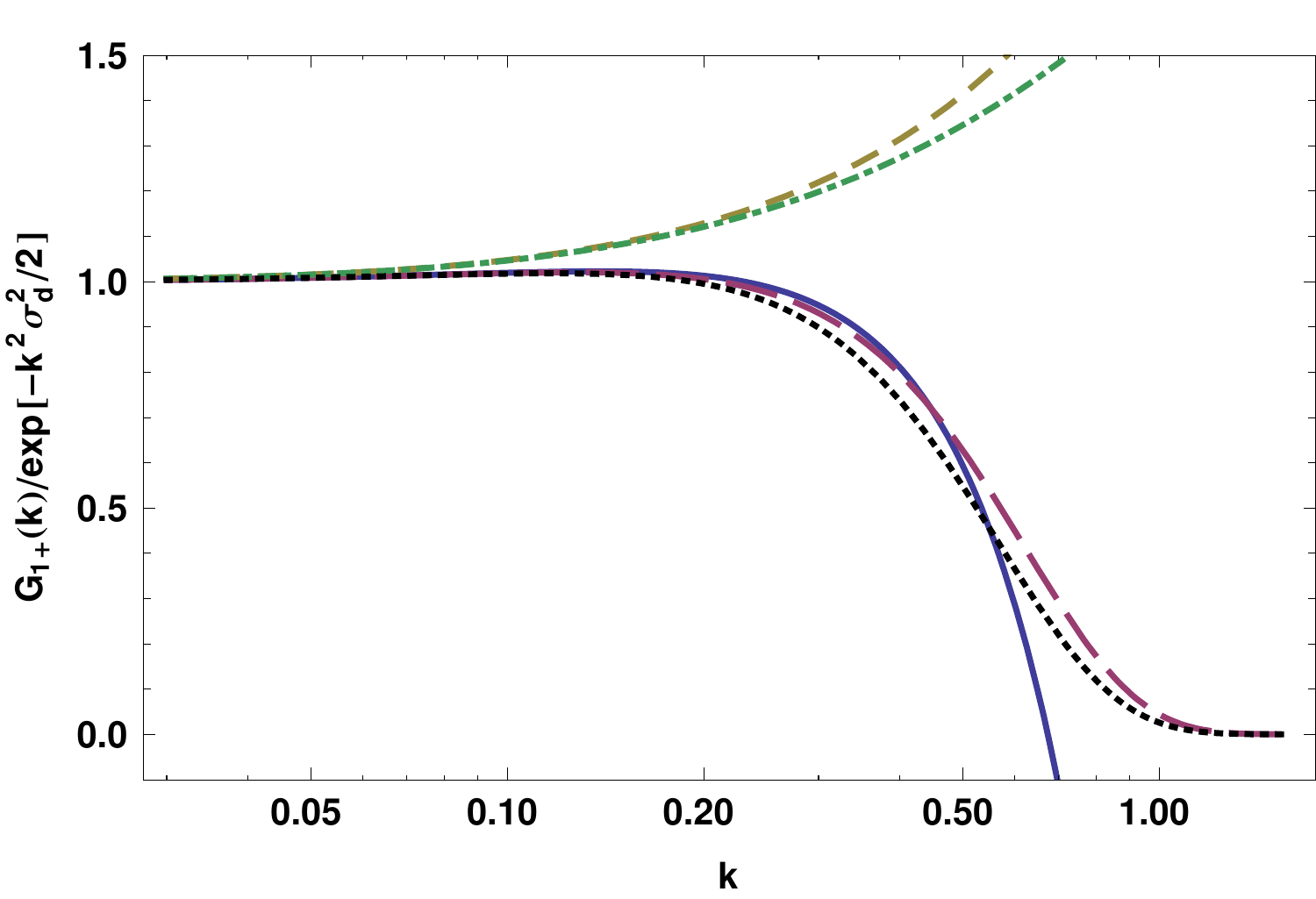}\hspace{.5cm}    
      \includegraphics[width=7cm]{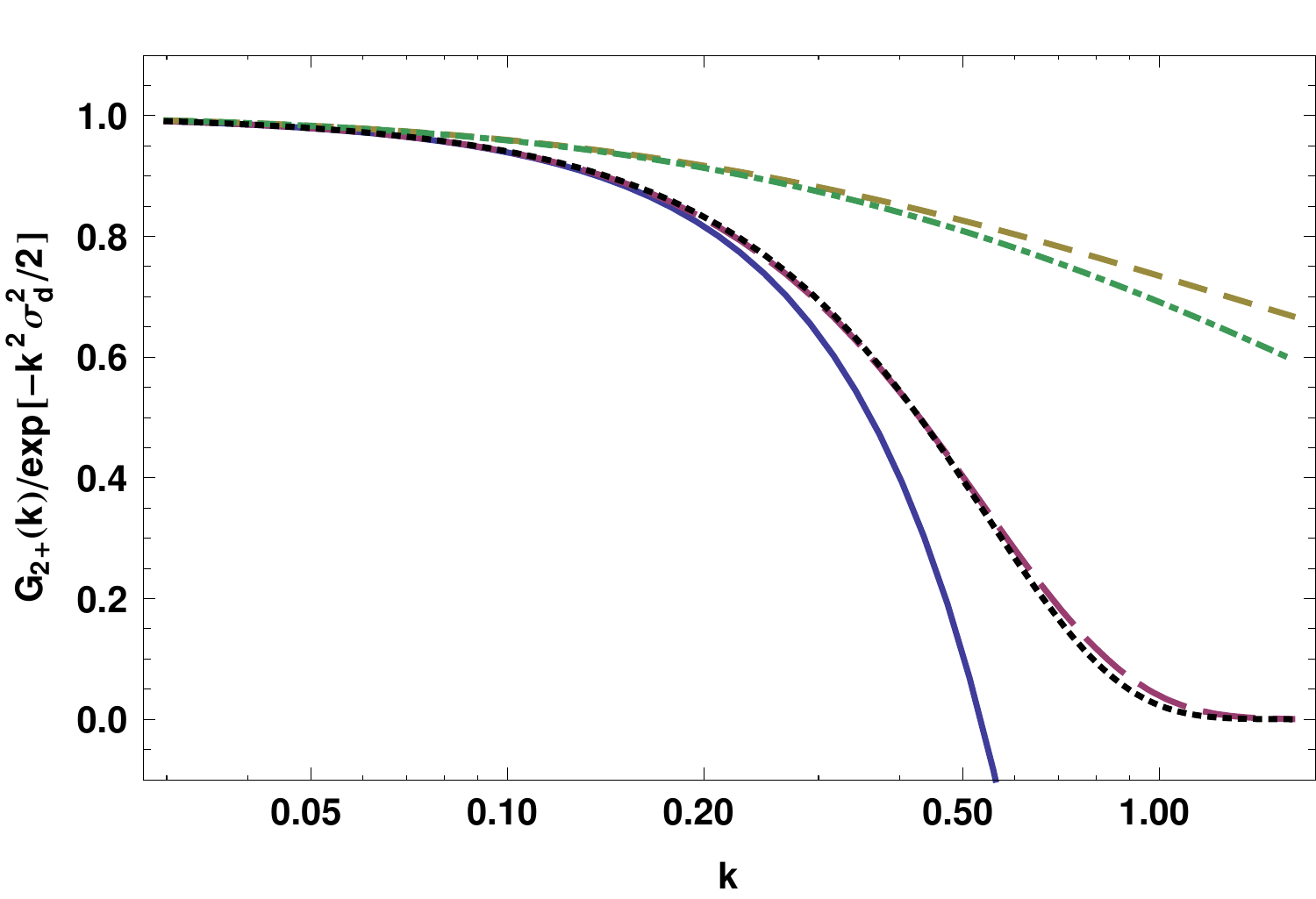} 
   \caption{The resulting propagators up to two loop order in units of the linear growing mode (top) and divided by the
   standard damping factor (bottom). Calculations are made for a standard $\Lambda$-CDM model at $z=0$. 
   The left panels correspond to the density field and the right panel to the reduced velocity divergence. 
The (yellow) dashed lines correspond to the RPT expression (at one loop order) and the dot-dashed lines (green) to the 1-loop regularized expression (\ref{GaplusReg1}), the (red) long dashed lines correspond to the two-loop RPT result (\ref{GaplusRPT2}), the (blue) solid lines to
the two-loop regularized expression (\ref{GaplusReg2}) and finally the black dotted lines to the alternative form of the two-loop regularized expression obtained from (\ref{regprop3}).}
   \label{propagators}
\end{figure}

We note however that it is however possible to build a RPT like expression, e.g. to exponentiate the two-loop result
and construct an alternative form that contains the same two-loop order terms,
At 1-loop order, the RPT propagator reads,
\begin{equation}
G_{a+}^{\RPT}(k,\eta)=e^{\eta}\,\exp\left(-\frac{1}{2}k^{2}\sigmad^{2}\,e^{2\eta}+\RegG_{a+}^{\oneloop}(k,\eta)\,e^{-\eta}\right).
\label{GaplusRPT}
\end{equation}
This expression can be extended into,
\begin{equation}
\begin{split}
G_{a+}^{\RPT-\twoloops}(k,\eta)=e^{\eta}\,\exp\left(
-\frac{1}{2}k^{2}\sigmad^{2}\,e^{2\eta}+\RegG_{a+}^{\oneloop}(k,\eta)\,e^{-\eta}\right.
\\
\left.-\frac{1}{2}\left[\RegG^{\oneloop}(k,\eta)\right]^{2}\,e^{-\eta}+\RegG_{a+}^{\twoloops}(k,\eta)\,e^{-\eta}\right).
\label{GaplusRPT2}
\end{split}
\end{equation}
This is a well-behaved expression since the second and third terms within the exponential is given by the second term of the right hand side of Eq. (\ref{alphafdef}) or (\ref{alphagdef}) and is negative. The resulting propagator is therefore always decreasing with $k$ and with time.
One can remark that if $\alpha_{f}(k,q_{1},q_{2})$ and $\alpha_{g}(k,q_{1},q_{2})$ vanish the expressions $G_{a+}^{\RPT}(k,\eta)$ and $G_{a+}^{\RPT-\twoloops}(k,\eta)$ are identical.

These various expressions are compared on Fig. \ref{propagators}. It appears that in practice, for a $\Lambda$-CDM model
the one-loop expressions, (\ref{GaplusRPT}) or (\ref{GaplusReg1}), lead to virtually indistinguishable results in the top panels. This is also to some extent the case for the two-loop results.  Differences between the one-loop and two-loop results are however significant. We will in the following compare those predictions with N-body results.

The bottom panels show more detailed comparisons of these predictions. It appears in particular that, unlike the RPT prediction at two-loop order, the $\RegG_{a+}^{\twoloops}(k,\eta)$ form can go negative at large enough $k$. This might be seen as a flaw (although this is of no consequences in practice because it takes place at scales that are quite damped) but it is corrected when one takes into account 
the one-loop correction to the displacement field (gray dotted lines) as it is the case when one uses Eq. (\ref{regprop3}). 

We finally give below the asymptotic forms of the multi-loop corrections in the low $k$ limit. Formally we have
\begin{equation}
G_{a+}(k,\eta)\approx e^{\eta}+c_{a}^{\oneloop}k^{2}\,e^{3\eta}+c_{a}^{\twoloops}k^{2}\,e^{5\eta}
\label{eq:k2asymp}
\end{equation}
where 
\begin{equation}
c_{1}^{\oneloop}=-4 \pi \frac{61}{630} \int\dd q\ P_{0}(q),\ \ 
c_{2}^{\oneloop}=- 4 \pi \frac{3}{13} \int\dd q\ P_{0}(q),\ \ 
\label{eq:k2asympv1}
\end{equation}
according to Eqs. (\ref{eq:fgasymlowk}) and
\begin{equation}
c_{a}^{\twoloops}=-(4\pi)^{2}\int\dd q_{1}\ \int\dd q_{2}\ \frac{q_{1}^{2}\ q_{2}^{2}}{q_{1}^{2}+q_{2}^{2}}\ \beta_{a}(q_{1},q_{2})\ P_{0}(q_{1})
\ P_{0}(q_{2}).
\label{eq:k2asympv2}
\end{equation}
More specifically, for the WMAP5 cosmological parameters we have,
\begin{eqnarray}
G_{1+}(k,\eta)&\approx& e^{\eta-\eta_{0}}
\left[1-\left(\frac{k}{0.31\ h \Mpc^{-1}}\right)^{2} e^{2\eta-2\eta_{0}}-\left(\frac{k}{0.60\ h\Mpc^{-1}}\right)^{2}e^{4\eta-4\eta_{0}}\right]\\
G_{2+}(k,\eta)&\approx& e^{\eta-\eta_{0}}
\left[1-\left(\frac{k}{0.17\ h \Mpc^{-1}}\right)^{2} e^{2\eta-2\eta_{0}}-\left(\frac{k}{0.89\ h\Mpc^{-1}}\right)^{2}e^{4\eta-4\eta_{0}}\right]
\label{eq:k2asympv3}
\end{eqnarray}
up to two-loop order 
where $\eta_{0}$ corresponds to $z=0$. We note that as scales in the quasi-linear regime -- say $k$ of the order of $0.1 h\Mpc^{-1}$ --
the two-loop corrections induce non-negligible effects irrespectively of the shape of the damping tail one observes at high $k$. 
This is clearly visible on the bottom panels of Fig. \ref{propagators}.

\begin{figure}[t] %  figure placement: here, top, bottom, or page
   \centering
{\hspace{-1.4cm}
\includegraphics[width=8.3cm]{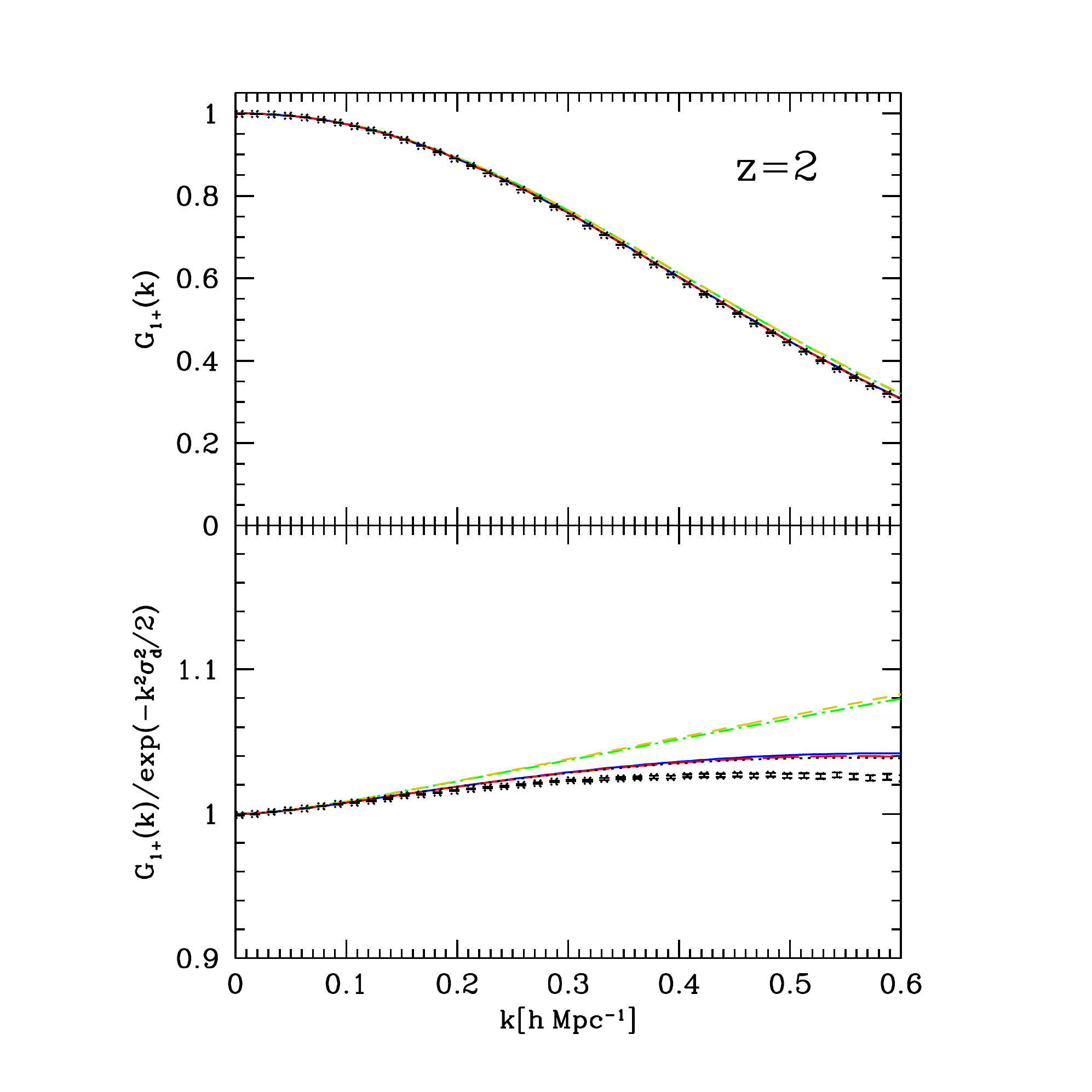} \hspace{-3cm}
\includegraphics[width=8.3cm]{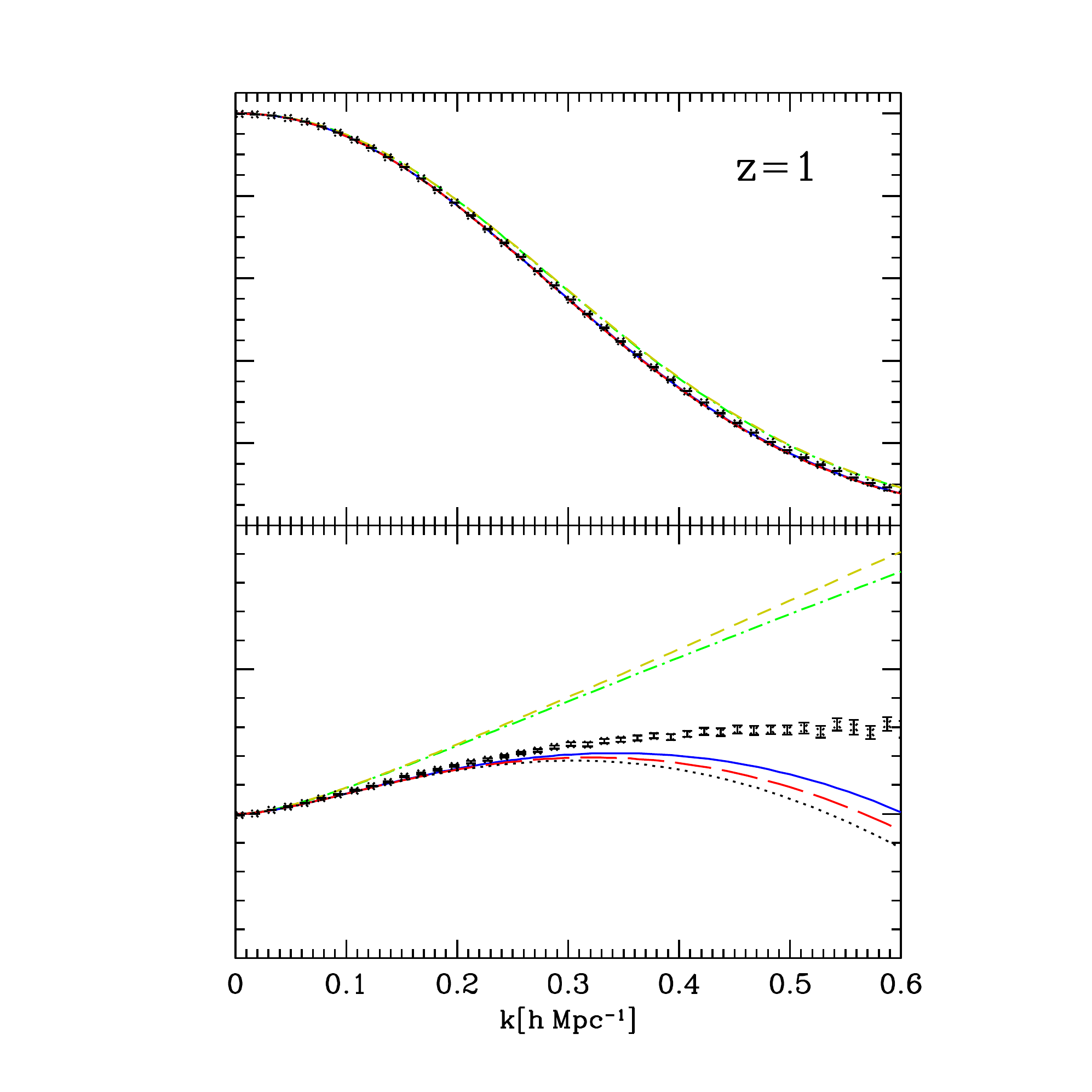}  \hspace{-3cm}
\includegraphics[width=8.3cm]{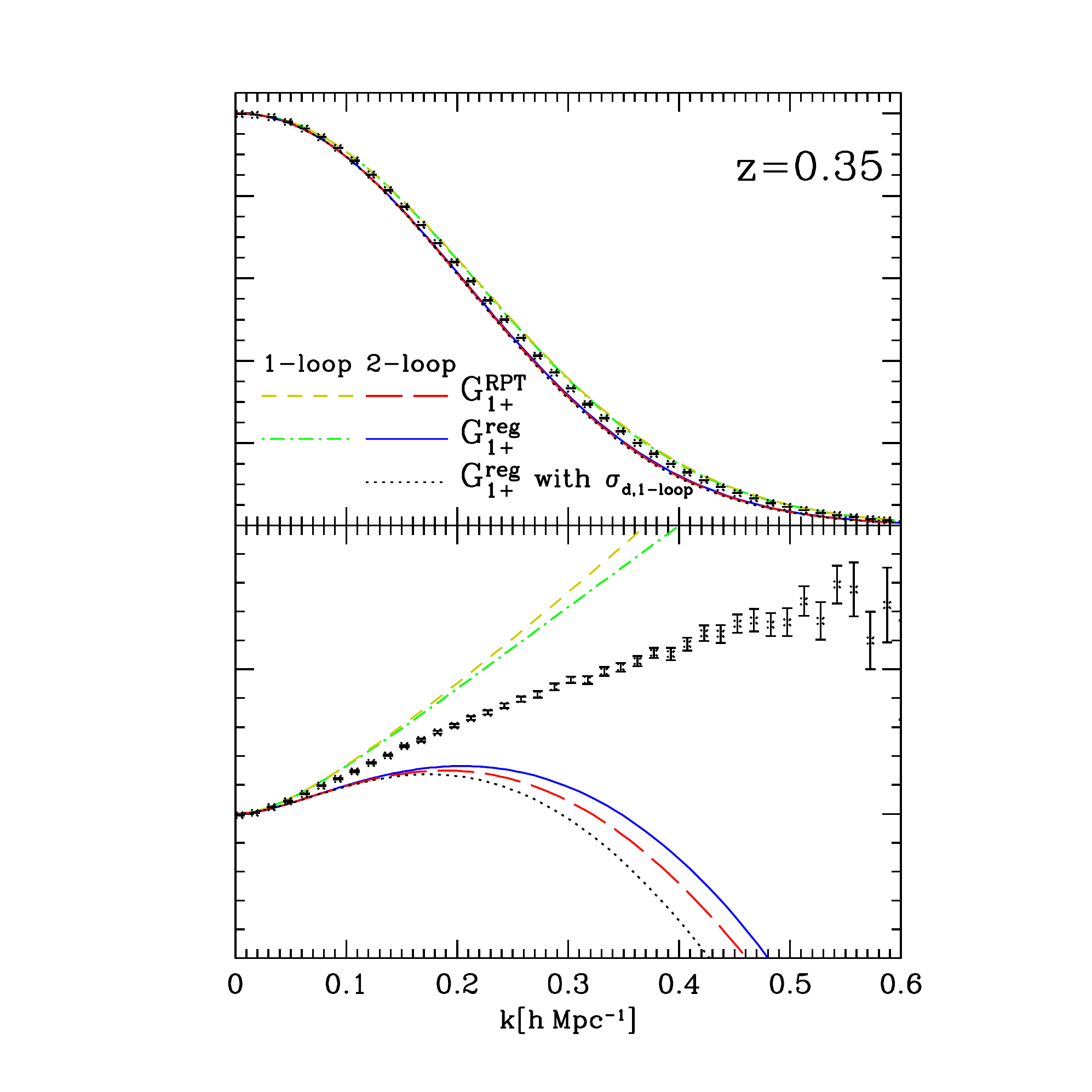} \hspace{-1.95cm}}
\vspace{-0.5cm}
   \caption{The regularized density propagator at 1 and 2-loop order compared to $N$-body results for $z=2$, $1$ and $0.35$. 
 The top panels correspond to the propagators in units of the linear growing factor  and the bottom panels to the propagators divided by the expected leading order damping factor.  The convention for the theoretical predictions are those of Fig. \ref{propagators}; the results from
 $N$-body simulations are shown with black points with error-bars.}
\label{NbodyComp}
\end{figure} 

\subsection{Comparisons with N-body results}

Finally we compare our results with N-body measurements. 
Here, we briefly explain the settings of the simulations as well as how we measured the propagators.

We have performed $N$-body simulations in periodic cubes with a side length of $2,048 h^{-1}\Mpc$
employing $N=1024^3$ particles. The initial conditions are created at $z=15$ using a parallel initial condition
generator developed in \cite{2009PASJ...61..321N,2011A&A...527A..87V}.
The code computes the displacement field based on the second-order Lagrangian perturbation theory 
(2LPT; \cite{1998MNRAS.299.1097S,2006MNRAS.373..369C}), and gives the displacements as well as the 
initial velocities to the particles placed on a regular lattice. 
In doing so, we adopt the best-fit flat $\Lambda$CDM cosmology to the five-year observations of WMAP
satellite \cite{Komatsu:2008hk}, and compute the linear matter power spectrum by {\tt CAMB} \cite{2000ApJ...538..473L}.
We created $60$ independent realizations of particle distributions, and simulate the gravitational evolution by a publicly 
available tree-PM code, {\tt Gadget2}. We store the outputs at $z=3, 2, 1$ and $0.35$.

We assign the particles onto $1024^3$ grid points using the Cloud-in-Cells interpolation scheme to obtain the final 
density field. This field is then Fourier transformed using the Fast Fourier transformations.
The propagators  are measured by taking the cross correlation between this field and the linear density field.
Note that here we do not use the initial particle distribution as the linear density. Instead, we employ the seed density
field which has been used as the source term in the 2LPT calculation. The former suffers from a slight nonlinearity,
while the latter is perfectly linear by construction.

We also perform simulations with different starting redshifts, box sizes, number of particles and internal parameters
which control the force accuracy to see the convergence of the propagators. We especially pay attention to
separating the effects of UV and IR cutoffs on the propagators, since we can cover only a finite wavenumber range
in simulations with finite number of particles. We made sure that our box size ($2,048 h^{-1}$ Mpc) is large enough to
avoid any systematic error larger than $1\%$ in the propagator originating from the IR cutoff. The UV cutoff, on the other
hand, can be a source of a small systematic decay of the propagator at small scale; this effect is almost independent on
the output redshift, and is typically $1\%$ at $k\sim0.3 h$ Mpc$^{-1}$, and $2\%$ at $k\sim0.4 h$ Mpc$^{-1}$.
The starting redshift, $z=15$, was chosen so that this systematic error is minimized given the box size and the number
of particles; the force error (the tree force, more precisely) damps structure and leads to a stronger decay of the 
propagator when started at a higher redshift, while the 2LPT is no longer applicable at lower redshift. 
See Nishimichi et al. (in prep.) for more details.

The propagator measured from the simulations is shown as symbols in Fig. \ref{NbodyComp} with error bars
showing the $1$-$\sigma$ uncertainties inferred from the variance among $60$ realizations.
We plot $G_{a+}(k)$ for the density field in the upper panels, and we also plot it after it is divided by
its high-$k$ asymptote to make a clearer comparison. At redshifts higher than $z=2$,
%%
%{\bf Comment by T.N.:  and $3$, if you want to include a panel for $z=3$), 
the 2-loop prediction is in good agreement with the simulation data.
The $N$-body data are slightly smaller than the two-loop prediction presumably due to the systematic error in the
$N$-body simulations (see above). On the other hand, both one-loop and two-loop predictions can not explain
the $N$-body data at lower redshifts ($z=0.35$ and and to a lesser extent $z=1$). 
The $N$-body data lie in between the two predictions. 
%%
%{\bf Comment by T.N.. You can freely change the taste on the level of agreement between sims and models.
%I do not intend to falsify your model based on the disagreement between sims and models.
%Also, please add some interpretation of the results, e.g., The 2-loop PT breaks down at $k=??$, or, 
%we do not care about the disagreement between the model and sims at $k=??$ where $\Gamma^{(1)}$ is already
%small, and $[\Gamma^{(1)}]^2(k)P_0(k)$ is only a subdominant contribution to the total power spectrum, e.t.c..}

%%%%%%%%%%%%%%%%%%%%%%%%%%%%%%%%%%%%%%%%%%%%%%%%%%%%%%%%%%%%%%%%%%%%%%%%%
%\section{Kernels of multi-loop propagators and convergence properties}
\section{Convergence properties}
\label{sec:multiloops}

In this section, we are interested in the sensitivity of the results to the IR and UV modes.
To this end, we introduce the notion of kernels that can be viewed as the response function of the propagator with respect to the initial linear spectrum.  This notion is also engrained in the approach proposed in 
\cite{2012arXiv1208.1191T} to obtain accelerated computations of the non-linear power spectra. 
We here first properly define these quantities and explore their small and large scale behaviors.

\subsection{Monte-Carlo integrations for the three-loop contribution}
\label{ThreeLoops}

\begin{figure}[t] %  figure placement: here, top, bottom, or page
   \centering
\includegraphics[width=10cm]{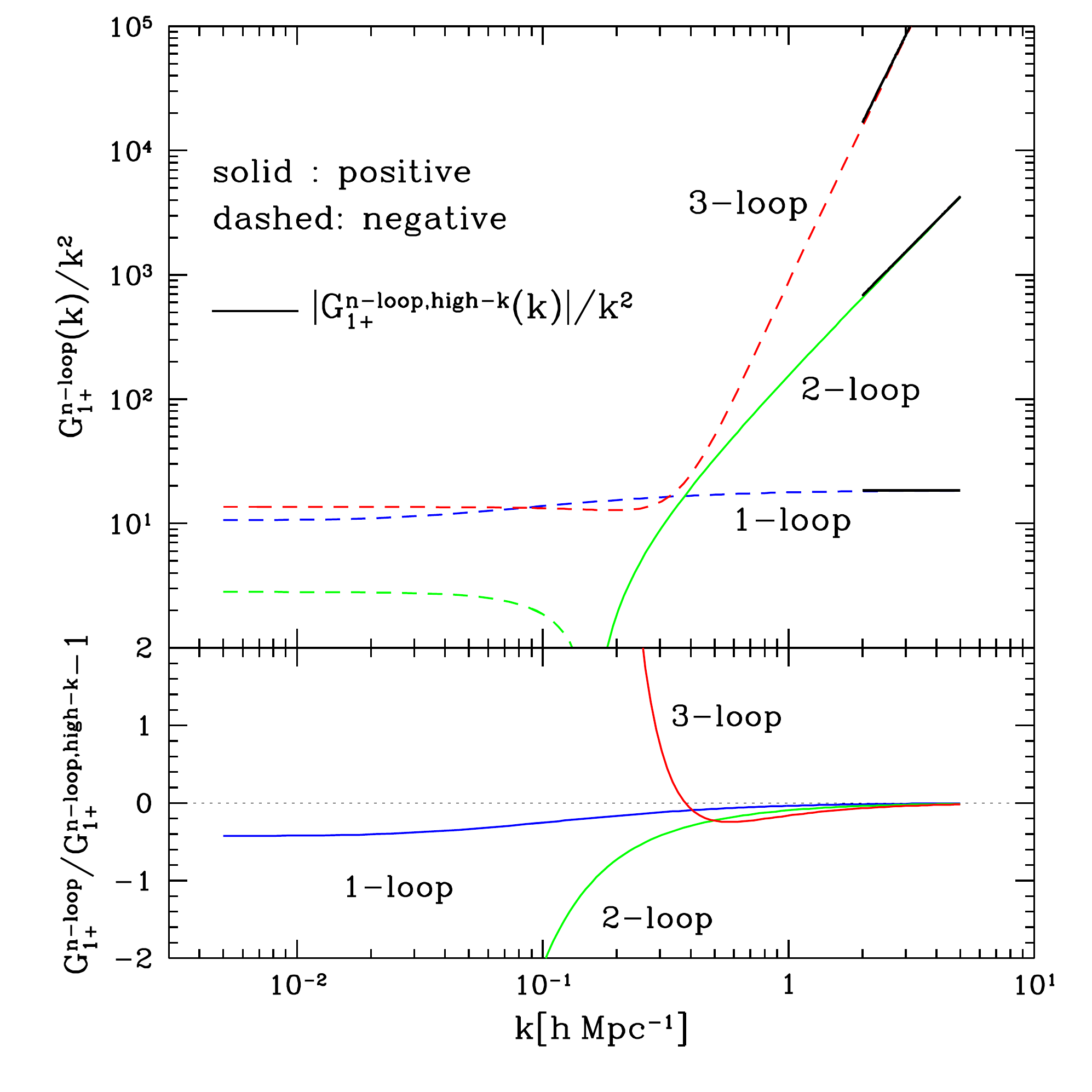} 
   \caption{Monte Carlo results for $G_{1+}^{\nloops}(k)$. The one-, two-, and three-loop corrections are respectively plotted as blue, green, and red lines. Top panel shows $G_{1+}^{\nloops}(k)/k^2$, while the bottom panel shows the fractional difference of $G_{1+}^{\nloops}(k)$ with its asymptotic behavior in the high $k$ limit $G_{1+}^{\nloops}(k)=(-k^{2}\sigmad^{2}/2)^{n}/n!$ (indicated by black solid lines in top panel).}
   \label{MCprop123}
\end{figure} 

\begin{figure}[t] %  figure placement: here, top, bottom, or page
   \centering
\includegraphics[width=9cm]{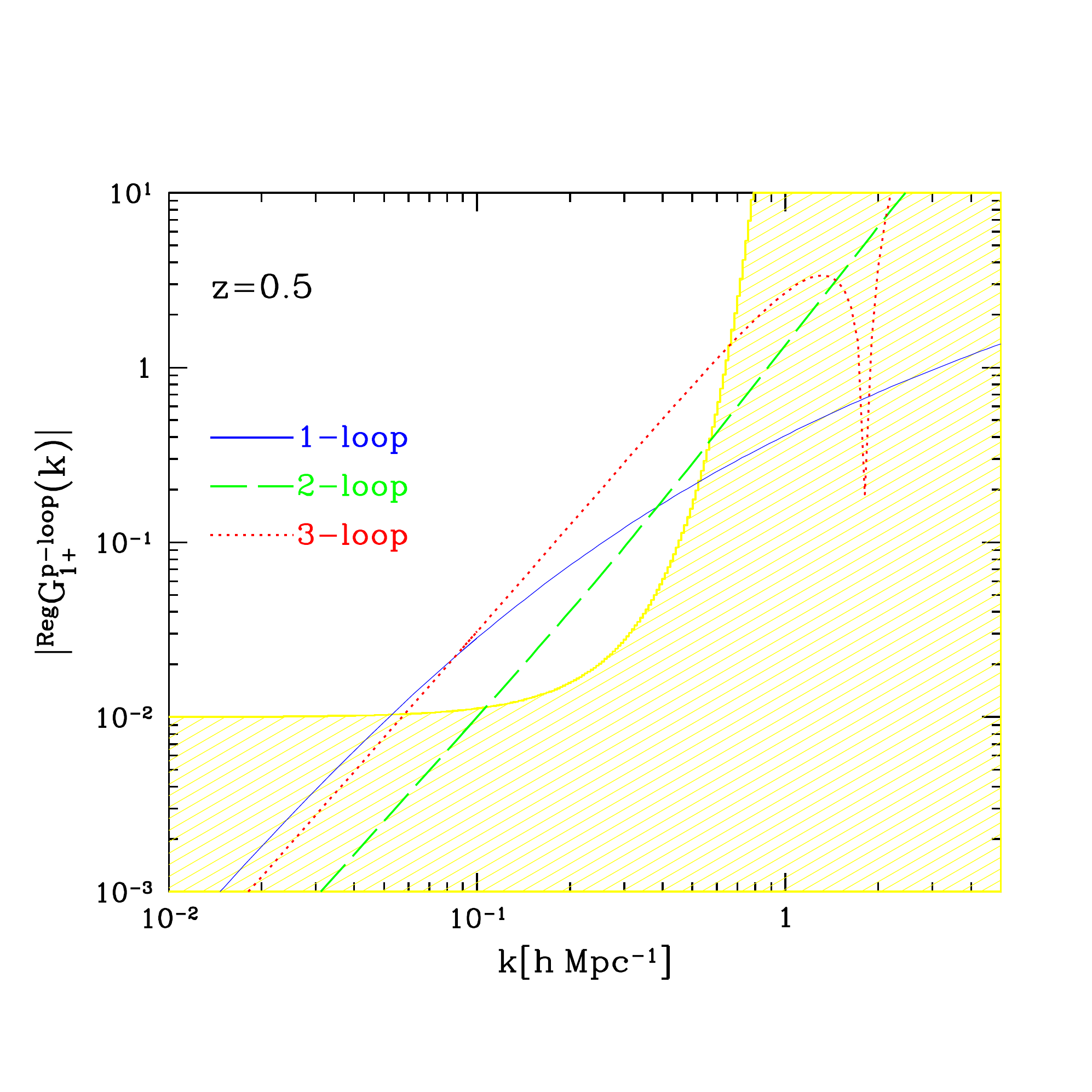} 
   \caption{Regular parts of the density propagator $\RegG_{1+}^{\ploops}(k)$ at one-, two-, and three-loop order with respectively solid, dashed and dotted lines.  The calculations are done for $z=0.5$. Note that each of this contribution scales with the redshift like $D_{+}(z)^{2p}$ where $p$ is the number of loops. The light yellow regions show the parameter space where the induced corrections to the \textsl{power spectrum} are less
than 1 percent. }
\label{RegG1plusz0}
\end{figure}

%\begin{figure}[t] %  figure placement: here, top, bottom, or page
%   \centering
%\includegraphics[width=9cm]{MCregprop123.pdf} 
%   \caption{Regularized propagator $\Gamma^{(1)}_{\reg}(k)$ at 1-, 2-, and 3-loop order (blue, green, and red). For comparison, standard PT
%propagators are also shown in dotted lines. Bottom: Same as in top panel, but we here plot $\Gamma^{(1)}_{\reg}(k)/\exp(-k^{2} \sigma_{d}^{2}/2)$.
%}
%\label{MCregprop123}
%\end{figure} 

In order to gain further insights into the properties of the loop corrections,
we have complemented our theoretical investigations with numerical evaluations of the three-loop contribution to the propagator. Unlike the two-loop contributions we did not try to obtain a semi-analytic expression of the results that could be used for any cosmological models. We rather aim here at evaluating the impact of the three-loop effects in a specific $\Lambda$-CDM like model. 
%(namely the model corresponding to the WMAP5 best fit parameters).

The results we present hereafter have been obtained from brute force calculations of $G_{1+}^{\threeloops}$, the two-point propagator for the 
density starting with growing mode initial conditions. 
The computations focus on the calculations of the density late time propagator at $z=0.5$ redshift. 
We apply the Monte-Carlo technique of quasi-random sampling using the library Cuba\footnote{http://www.feynarts.de/cuba/}, and directly perform $2D$, $5D$ and $8D$ integrations for $G_{1+}^{\oneloop}$, $G_{1+}^{\twoloops}$ and $G_{1+}^{\threeloops}$ respectively. The transfer function of the linear power spectrum is calculated with CAMB code adopting the cosmological parameters determined by WMAP5 \cite{Komatsu:2008hk}.

In Fig. \ref{MCprop123} we separately show the one-, two-, and three-loop corrections to the two-point propagator (blue, green, and red,
respectively). In top panel, to reduce the dynamic range, the results are divided by $k^2$, i.e., $G_{1+}^{\nloops}(k)/k^{2}$, and are plotted as function of wavenumber in logarithmic scales. Note that the solid and dashed lines respectively indicate the
positive and negative values in amplitude. The Monte Carlo results for one- and two-loop corrections accurately reproduce
the analytic calculations. At low-k regime, all the plotted results show scale-independent behavior, indicating that
the corrections behave like  $G_{1+}^{\nloops}(k) \sim k^{2}$. On the other hand, at high-k regime, the three curves behave
according to the expected forms described by Eq. (\ref{ploopsd}).  They tend to approach the asymptotic form, $G_{1+}^{\nloops}(k) = (-k^{2}\sigmad^{2}/2)^n/n!$, depicted as black solid lines. To see this clearly, the bottom panel of Fig. \ref{MCprop123} shows the fractional difference between $G_{1+}^{\nloops}(k)$ and its asymptotic behavior, i.e., $G_{1+}^{\nloops}(k)/ \left[(-k^{2}\sigmad^{2}/2)^n/n!\right]-1$. At $k\ge1 h \Mpc^{{-1}}$,  all the three curves consistently converge to zero. We can actually verified that the regular part of each contribution is finite at large $k$.

In Fig. \ref{RegG1plusz0} we show the relative importance of the loop corrections as a function of scale. What is plotted are  the regular parts of the terms as defined in
subsection \ref{sec:regparts}. It appears that at $k=0.1 h/\Mpc$ and at $z=0.5$ the one-loop contribution introduces a 5\% correction, the two-loop contribution a further 2\% contribution. At $k=0.2  h/\Mpc$ the two contributions are both at 10\% level but then the mode damping is already significant. Note that as the redshift is increasing the importance of all these loop corrections decreases, their magnitude scaling like $D_{+}^{2p}$ where $D_{+}$ is the linear growth rate and $p$ is the number of loops. 

Obviously, we hoped and expected the three-loop contribution to be smaller that the two  other terms for $k\le 1h/\Mpc$ but this is not the case. At $k=0.1 h/\Mpc$ it gives a 10\% effect! To say the least this suggests that the three-loop corrections, assuming they are correctly evaluated by our Monte-Carlo integrations, cannot help to match the N-body results at low redshift. Given the results shown on Fig. \ref{MCprop123} we have not much reasons to doubt the validity of the numerical integration. To better apprehend the origin of these effects we hereafter introduce the notion of kernels.

%Summing up each contribution to the one-point propagator, Fig. \ref{MCregprop123} shows the regularized propagator at 1-, 2-, and 3-loop order (blue, green, and red), respectively following Eq. (\ref{regprop2}).  The top panel of Fig. \ref{MCregprop123} also shows the results for the standard PT propagator. The standard PT results eventually diverge at higher-k, while the regularized propagator gives a finite
%amplitude and rapidly decays at $k\le 0.6h\,\Mpc^{-1}$. However, the regularized PT result at 3-loop order is somewhat
%strange, and it becomes negative at $0.2 h \,\Mpc^{-1}\le k\le 0.7 h \,\Mpc^{-1}$. This is because the 3-loop correction, whose
%amplitude is negative and is comparable to the one of the 1-loop correction (see top panel of Fig. \ref{MCprop123}), gives a rather
%large contribution to the intermediate scales. The significance of the 3-loop correction is also seen in the bottom panel
%of Fig. \ref{MCregprop123}, where the regularized propagators divided by the exponential factor, $\Gamma^{(1)}_{\reg}(k)/\exp(-k^{2}\sigmad^{2}/2) $are plotted. A large reduction of $\Gamma^{(1)}_{\reg} (k)$ is manifest for the result at 3-loop order. 

%{\bf Notes from Atsushi to be included + figs}
%{\bf To be added also is shape of kernel plus dependence of the results with cut-off}

\subsection{Definitions and properties of the kernel functions}

Let us define the functions $K^{\ploops}(k,q)$ as the kernels from which the contribution of the two-point propagator to $p$-loop order can be reconstructed,
\begin{equation}
\RegG_{a+}^{\ploops}(k)=\int \frac{\dd q}{q}\ K_{a+}^{\ploops}(k,q)P_{0}(q).
%K_{a+}^{\ploops}(k,q)=\frac{\delta \RegG_{a+}^{\ploops}(k)}{\delta P_{0}(q)}
\end{equation}
We have then for instance
\begin{eqnarray}
K_{1+}^{\oneloop}(k,q)&=&4\pi q^{3}\left(f(q,k)+\frac{1}{6}\frac{k^{2}}{q^{2}}\right)\\
K_{1+}^{\twoloops}(k,q)&=&-(4\pi)^{2}\,q^{3}\int{\dd q_{1}}\,\frac{q_{1}^{2}\,k^{2}}{q_{1}^{2}+q^{2}}\,
\alpha_{f}\left(\frac{q_{1}}{k},\frac{q}{k}\right)\,P_{0}(q_{1}).
\end{eqnarray}
Note that the kernel functions depend themselves a priori on the initial power spectrum: $K_{a+}^{\oneloop}(k,q)$
is a tree order object, $K_{a+}^{\twoloops}(k,q)$ a one-loop order object (and therefore a linear function of $P_{0}(q)$), etc.
These functions give, for each order, the impact of a linear mode $q$ on the amplitude of the late time mode $k$ we are interested in. In particular it tells how the small-scale modes affect the large-scale modes under consideration.  In the following we will focus our interest in understanding the 
high-$q$ behavior of the kernel functions $K(k,q)$.

\begin{figure}[ht] %  figure placement: here, top, bottom, or page
   \centering
   \includegraphics[width=10cm]{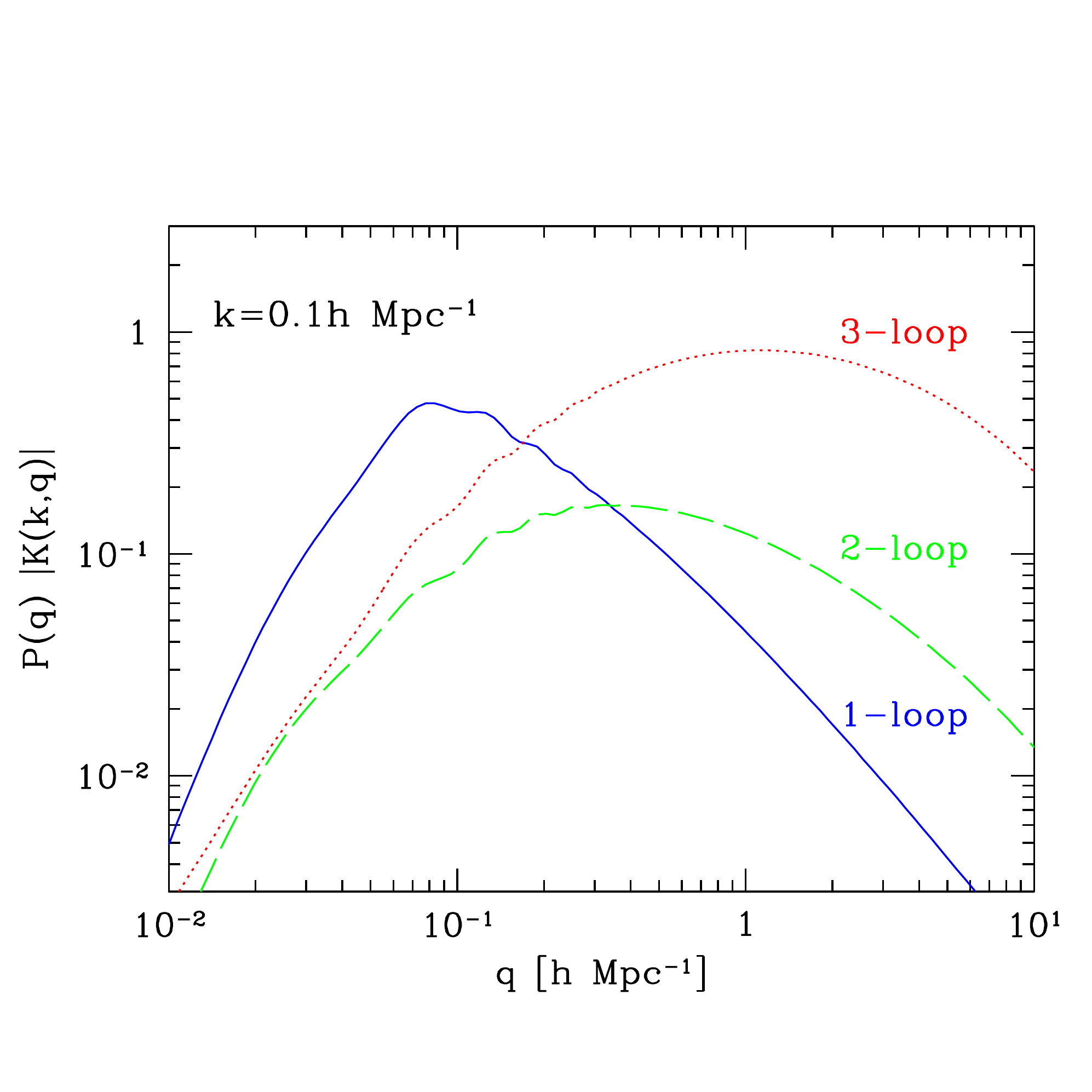}
   \caption{The shape of the kernel functions $P_{0}(q)K^{\oneloop}(k,q)$ (blue solid line), $P_{0}(q)K^{\twoloops}(k,q)$ (green dashed line) for $k=0.1h/\Mpc$ and $P_{0}(q)K^{\threeloops}(k,q)$ (red dotted line)  as a function of $q$ for $z=0.5$.}
   \label{Kernels}
\end{figure}

In Fig. \ref{Kernels} we show the shape of the kernel functions at one,  two-loop and three-loop order for $k=0.1 h/\Mpc$.  The dashed line corresponds to the one-loop expression. As can be seen it is rather peaked at $q\approx k$ and we have
\begin{eqnarray}
K_{1+}^{\oneloop}(k,q)P_{0}(q)&=&\frac{464\pi}{315}q^{3}P_{0}(q)\ \ {\rm for}\ q\ll k\\
K_{1+}^{\oneloop}(k,q)P_{0}(q)&=&\frac{176 \pi}{315}\,k^{2}\, q\ P(q)\ \ {\rm for}\ q\gg k\label{K1plusAsymp}
\end{eqnarray}
At two-loop order, the behaviors are qualitatively different. The function peaks rather for $q=0.5h/\Mpc$, \textsl{irrespectively} of the value for $k$
(when $k<0.5h/\Mpc$). We note that
\begin{equation}
K_{1+}^{\twoloops}(k,q)P_{0}(q)\sim k^{2} q^{2} P_{0}(q)\ \ {\rm for}\ q\gg k\label{K2plusAsymp}
\end{equation}
so that the convergence is obtained for a spectral index smaller than $-2$. This corresponds to the result mentioned in the beginning of subsection
\ref{subsec:resprop}.
These trends are amplified for the three-loop results shown with a dot-dashed line for which an even lower power law index is required 
for convergence. In general the convergence properties of the multi-loop kernel are determined by the properties of the functions $F_{n}(\vq_{i})$ and
$G_{n}(\vq_{i})$ and how they behave when one or their argument is, in norm, much larger than the sum of the wave modes.
As mentioned in  \cite{Scoccimarro:1996se}
it is to be noted that the Galilean invariance of the motion equation implies that,
\begin{equation}
F_{n}(\vq_{1},\dots,\vq_{n})\sim \frac{\vert \sum_{j}q_{j}\vert^{2}}{q_{i}^{2}}\ \ {\rm when}\ q_{i}\gg \vert \sum_{j}q_{j}\vert,
\label{IRasymp}
\end{equation}
whenever one of the $q_{i}$ is much larger than the sum. This can be seen at an elementary level on the properties of the vertex function
$\alpha(\vk_{1},\vk_{2})$ and $\beta(\vk_{1},\vk_{2})$: they both vanish when the sum of the argument goes to 0.
The property (\ref{IRasymp}) has direct consequences on the properties of the loop corrections. As a result,
the $p-$loop correction takes indeed the form,
\begin{equation}
G^{\ploops}(k,\eta)=(4\pi)^{p}\int q_{1}^{2}\dd q_{1}\,\dots\, q_{p}^{2}\dd q_{p}\ P_{0}(q_{1})\dots P_{0}(q_{p})\,\mF^{\ploops}(k,q_{1},\dots,q_{p})
\end{equation}
with
\begin{equation}
\mF^{\ploops}(k,q_{1},\dots,q_{p})=\frac{(2p+1)!!}{(4\pi)^{p}}\int\dd^{2}\hq_{1}\dots \dd^{2}\hq_{p}\ F_{2p+1}(\vk,\vq_{1},-\vq_{1},\dots,\vq_{p},-\vq_{p})
\end{equation}
where the integration is here restricted to the wave mode angles. The property (\ref{IRasymp}) then implies that 
\begin{equation}
\mF^{\ploops}(k,q_{1},\dots,q_{p})\sim \frac{k^{2}}{q_{i}^{2}}\ \ {\rm when}\ q_{i}\gg k.
\end{equation}
A simple and effective way to capture this property is to write $\mF(k,q_{1},\dots,q_{p})$ following the functional form,
\begin{equation}
\mF(k,q_{1},\dots,q_{p})=\frac{k^{2}}{\sum_{i }q_{i}^{2}}\alpha_{p}(k/q_{1},\dots, k/q_{p})
\end{equation}
generalizing Eqs. (\ref{alphafdef},\ref{alphagdef}) and where $\alpha(k/q_{1},\dots,k/q_{p})$ is always finite. This property determines the converging properties of
the $p-$loop correction to the propagator. Clearly the converging properties of the loop corrections deteriorate. It can easily be shown that formally, for a power law spectrum of index $n_{s}$, the convergence is ensured only when,
\begin{equation}
n_{s}<-3+\frac{2}{p}.
\end{equation}
It clearly appears that the convergence domains shrink with the order. It leads to a greater sensitivity to the large wave-modes
as one increases the loop order. What we stress here is that the kernel functions can explicitly be computed and thus one can 
exactly quantify the impact of large wave modes to each $p$-loop correction terms.

Note that it is possible to define in general the kernel of contributing terms as
\begin{equation}
\delta P_{ab}^{\sharp}(k)=\int \frac{\dd q}{q}\ K_{ab}^{\sharp}(k,q)P_{0}(q)
\end{equation}
where $\sharp$ denotes a peculiar term. One can then compare the kernels of the different terms that contribute to the non-linear power spectrum.
It is worth mentioning that the kernels for the two-point propagators are the broadest. Indeed any other kernel function in the large-$q$ limit 
is going to take advantage of the asymptotic form (\ref{IRasymp}) at least two times instead of one so that
\begin{equation}
K_{1+}^{\sharp}(k,q)P_{0}(q)\sim \frac{k^{4}}{q^{4}}\ q^{3l} P_{0}(q)\ \ {\rm for}\ q\gg k
\label{KnpAsymp}
\end{equation}
where $l$ is the number of loops appearing in the diagram $\sharp$. This is to be compared with Eqs. (\ref{K1plusAsymp}, \ref{K2plusAsymp}).

Clearly the loop corrections to the propagators are then  the contributions built from perturbation theory calculations that are most sensitive 
to the small scale modes. At low redshift our findings imply that a regularization procedure in the UV domain should be used in order to make the contributions of these terms realistic. Unless partial resummation of higher order terms could provide us with a self-consistent regularization 
scheme, it should be derived from external pieces of information. This regularization could be associated for instance from the development 
multi-flow regimes. For instance it was advocated in  \cite{2012arXiv1206.2926C} 
that the impact of the small scale physics could be captured with an effective theory 
approach where effective anisotropic pressure terms are introduced in the fluid. Our results however suggest that such effects are
subdominant at the one-loop order - the contributions of such terms need not be recalculated because they are naturally protected 
enough from UV physics - but not at two-loop  order for $z\le 0.5$. It also suggests that when $z$ gets larger the importance of such corrections should vanish away even at 2 loop order at the nonlinear regime starts at larger wave-modes.

\section{Conclusions}
\label{sec:conclusions}

Two-point propagators are key ingredients for the computation of power spectra in a cosmological context. These are also quantities for which it is easier to make analytical predictions with a Perturbation Theory approach in parts because they
are expected to be damped in the large-wave modes limit in parts because the number of terms in the loop corrections  are more easily computable.  In this paper we have presented the most advanced results describing the expected behavior of the two-point propagators from a Perturbation Theory point of view. In particular we have exploited and extented the forms presented in \cite{2011arXiv1112.3895B} to propose theoretical expressions for the two-point propagators that incorporate up to two-loop results. The examination of these forms led to the concept of regular parts for loop contributions that are found to be finite even when the r.m.s. of the displacement field, $\sigmad$, goes to infinity: regular 
parts are found to be the parts that contribute to the expression of the propagators once the dependence with $\sigmad$ has been 
explicitly taken into account. They also can be used as a consistency test for Monte-Carlo numerical calculations of the propagators 
as they predict the leading $k$ behavior of the contributing terms. 

The main part of this paper is the explicit calculation of the two-loop contributions to the two-point propagator. We were not able to produce a simple analytical form  but we rather present, in section \ref{sec:thediagrams},  simple fitting functions that capture all the expected properties of these contributions, in particular their asymptotic properties. 
These forms can be used for any cosmological models.  Note that we propose different variants for the two-loop expression of the two-point propagator but they differ only mildly. There results are presented in Fig. \ref{propagators}. This is the main result of the paper. 
It has been compared with $N$-body results. They show that two-loop corrections help significantly in predicting the shape of the two-point density propagators at $z\ge 0.5$. 

For lower redshifts the two-loop terms  appear to be too large  when compared to $N$-body codes. We found that a possible reason for such a discrepancy is due to the increasing sensitivity of the loop corrections to the small-scale modes. A significant fraction of the contributions indeed originate  from modes  that are in the fully nonlinear regime. At those scales the whole perturbation theory scheme breaks down as 
the fluid has not only evolved into the nonlinear regime but also contains multi flow regions with order unity anisotropic pressure. The latter is not captured by the motion equations we use and should induce extra contributions as it has been put forward in recent papers,  \cite{2012arXiv1206.2926C,PieManSav1108}. We do not think however that
it has significant impact on the one-loop results. Those effects are rather expected to be significant at two-loop order and for $z\le 0.5$. 
This sensitivity to the small scale physics is shown to be stronger for the three-loop contribution that we computed with the help of 
Monte-Carlo integrations.

About the converging properties of the Perturbation Theory series we note that loop corrections, at any order, induce large-scale $k^2$ corrections to the propagators and consequently to the power spectra. They are explicitly given in Eqs. (\ref{eq:k2asymp}-\ref{eq:k2asympv3}) up to two-loop order
with coefficients that depends on the overall shape on the linear power spectrum.
Again it stresses the relative importance of the sensitivity  of the standard Perturbation theory calculation to the UV domain. 
We also note that this
sensitivity is controlled by the Galilean invariance which leads to the general property given in Eq.  (\ref{IRasymp}). This form appears to be a key ingredient  for the converging
properties of the Perturbation Theory series. It indeed sets the converging properties of both the diagrams that correspond to propagator loop corrections, 
as given by Eqs. (\ref{K1plusAsymp}, \ref{K2plusAsymp}),
and the other ones,  as depicted in Eq. (\ref{KnpAsymp}).
Clearly if this property is dropped, the sensitivity of loop corrections to the UV domain  is more severe. We note that if in standard gravity cosmologies, the relation (\ref{IRasymp}) is satisfied, this is not necessarily so in cosmological models incorporating modified gravity effects or in general a clustering quintessence component. This is the case for instance in models explored in \cite{2011JCAP...06..019B,2009PhRvD..80j4005C,2012arXiv1201.3614C} that incorporate extra dynamical degrees of freedom~\footnote{This is not the case however in  \cite{2011JCAP...03..047S} in the limit that has been chosen (as indeed the coupling function $\alpha(\vk_{1},\vk_{2})$ and $\beta(\vk_{1},\vk_{2})$ are left unchanged)}.

\begin{acknowledgments}
Numerical computations for the present work have been carried out in part
on Cray XT4 at Center for Computational Astrophysics,
CfCA, of National Astronomical Observatory of Japan,
and in part under the Interdisciplinary Computational
Science Program in Center for Computational Sciences,
University of Tsukuba.
\\
This work has been benefited from exchange visits supported by a 
bilateral grant from 
Minist\`ere  Affaires Etrang\`eres et Europ\'eennes in France
and Japan Society for the Promotion of Science (JSPS).  
A.T. acknowledges support from Institutional Program for Young
Researcher Overseas Visit funded by the JSPS. 
A.T. is also supported in part by a Grant-in-Aid for Scientific 
Research from the JSPS (No.~24540257). 
T.N. is supported from the JSPS.  F.B is also partly supported by the French Programme National de Cosmologie et Galaxies.
\end{acknowledgments}

\bibliography{LSStructure}

\begin{thebibliography}{35}
\expandafter\ifx\csname natexlab\endcsname\relax\def\natexlab#1{#1}\fi
\expandafter\ifx\csname bibnamefont\endcsname\relax
  \def\bibnamefont#1{#1}\fi
\expandafter\ifx\csname bibfnamefont\endcsname\relax
  \def\bibfnamefont#1{#1}\fi
\expandafter\ifx\csname citenamefont\endcsname\relax
  \def\citenamefont#1{#1}\fi
\expandafter\ifx\csname url\endcsname\relax
  \def\url#1{\texttt{#1}}\fi
\expandafter\ifx\csname urlprefix\endcsname\relax\def\urlprefix{URL }\fi
\providecommand{\bibinfo}[2]{#2}
\providecommand{\eprint}[2][]{\url{#2}}

\bibitem[{\citenamefont{{Bernardeau} et~al.}(2011)\citenamefont{{Bernardeau},
  {Crocce}, and {Scoccimarro}}}]{2011arXiv1112.3895B}
\bibinfo{author}{\bibfnamefont{F.}~\bibnamefont{{Bernardeau}}},
  \bibinfo{author}{\bibfnamefont{M.}~\bibnamefont{{Crocce}}}, \bibnamefont{and}
  \bibinfo{author}{\bibfnamefont{R.}~\bibnamefont{{Scoccimarro}}},
  \bibinfo{journal}{ArXiv e-prints}  (\bibinfo{year}{2011}),
  \eprint{1112.3895}.

\bibitem[{\citenamefont{{Eisenstein} et~al.}(2005)\citenamefont{{Eisenstein},
  {Zehavi}, {Hogg}, {Scoccimarro}, {Blanton}, {Nichol}, {Scranton}, {Seo},
  {Tegmark}, {Zheng} et~al.}}]{2005ApJ...633..560E}
\bibinfo{author}{\bibfnamefont{D.~J.} \bibnamefont{{Eisenstein}}},
  \bibinfo{author}{\bibfnamefont{I.}~\bibnamefont{{Zehavi}}},
  \bibinfo{author}{\bibfnamefont{D.~W.} \bibnamefont{{Hogg}}},
  \bibinfo{author}{\bibfnamefont{R.}~\bibnamefont{{Scoccimarro}}},
  \bibinfo{author}{\bibfnamefont{M.~R.} \bibnamefont{{Blanton}}},
  \bibinfo{author}{\bibfnamefont{R.~C.} \bibnamefont{{Nichol}}},
  \bibinfo{author}{\bibfnamefont{R.}~\bibnamefont{{Scranton}}},
  \bibinfo{author}{\bibfnamefont{H.-J.} \bibnamefont{{Seo}}},
  \bibinfo{author}{\bibfnamefont{M.}~\bibnamefont{{Tegmark}}},
  \bibinfo{author}{\bibfnamefont{Z.}~\bibnamefont{{Zheng}}},
  \bibnamefont{et~al.}, \bibinfo{journal}{\apj} \textbf{\bibinfo{volume}{633}},
  \bibinfo{pages}{560} (\bibinfo{year}{2005}), \eprint{arXiv:astro-ph/0501171}.

\bibitem[{\citenamefont{{Percival} et~al.}(2007)\citenamefont{{Percival},
  {Nichol}, {Eisenstein}, {Weinberg}, {Fukugita}, {Pope}, {Schneider},
  {Szalay}, {Vogeley}, {Zehavi} et~al.}}]{2007ApJ...657...51P}
\bibinfo{author}{\bibfnamefont{W.~J.} \bibnamefont{{Percival}}},
  \bibinfo{author}{\bibfnamefont{R.~C.} \bibnamefont{{Nichol}}},
  \bibinfo{author}{\bibfnamefont{D.~J.} \bibnamefont{{Eisenstein}}},
  \bibinfo{author}{\bibfnamefont{D.~H.} \bibnamefont{{Weinberg}}},
  \bibinfo{author}{\bibfnamefont{M.}~\bibnamefont{{Fukugita}}},
  \bibinfo{author}{\bibfnamefont{A.~C.} \bibnamefont{{Pope}}},
  \bibinfo{author}{\bibfnamefont{D.~P.} \bibnamefont{{Schneider}}},
  \bibinfo{author}{\bibfnamefont{A.~S.} \bibnamefont{{Szalay}}},
  \bibinfo{author}{\bibfnamefont{M.~S.} \bibnamefont{{Vogeley}}},
  \bibinfo{author}{\bibfnamefont{I.}~\bibnamefont{{Zehavi}}},
  \bibnamefont{et~al.}, \bibinfo{journal}{\apj} \textbf{\bibinfo{volume}{657}},
  \bibinfo{pages}{51} (\bibinfo{year}{2007}), \eprint{arXiv:astro-ph/0608635}.

\bibitem[{\citenamefont{{Blake} et~al.}(2011)\citenamefont{{Blake}, {Davis},
  {Poole}, {Parkinson}, {Brough}, {Colless}, {Contreras}, {Couch}, {Croom},
  {Drinkwater} et~al.}}]{2011MNRAS.415.2892B}
\bibinfo{author}{\bibfnamefont{C.}~\bibnamefont{{Blake}}},
  \bibinfo{author}{\bibfnamefont{T.}~\bibnamefont{{Davis}}},
  \bibinfo{author}{\bibfnamefont{G.~B.} \bibnamefont{{Poole}}},
  \bibinfo{author}{\bibfnamefont{D.}~\bibnamefont{{Parkinson}}},
  \bibinfo{author}{\bibfnamefont{S.}~\bibnamefont{{Brough}}},
  \bibinfo{author}{\bibfnamefont{M.}~\bibnamefont{{Colless}}},
  \bibinfo{author}{\bibfnamefont{C.}~\bibnamefont{{Contreras}}},
  \bibinfo{author}{\bibfnamefont{W.}~\bibnamefont{{Couch}}},
  \bibinfo{author}{\bibfnamefont{S.}~\bibnamefont{{Croom}}},
  \bibinfo{author}{\bibfnamefont{M.~J.} \bibnamefont{{Drinkwater}}},
  \bibnamefont{et~al.}, \bibinfo{journal}{\mnras}
  \textbf{\bibinfo{volume}{415}}, \bibinfo{pages}{2892} (\bibinfo{year}{2011}),
  \eprint{1105.2862}.

\bibitem[{\citenamefont{{Nuza} et~al.}(2012)\citenamefont{{Nuza}, {Sanchez},
  {Prada}, {Klypin}, {Schlegel}, {Gottloeber}, {Montero-Dorta}, {Manera},
  {McBride}, {Ross} et~al.}}]{2012arXiv1202.6057N}
\bibinfo{author}{\bibfnamefont{S.~E.} \bibnamefont{{Nuza}}},
  \bibinfo{author}{\bibfnamefont{A.~G.} \bibnamefont{{Sanchez}}},
  \bibinfo{author}{\bibfnamefont{F.}~\bibnamefont{{Prada}}},
  \bibinfo{author}{\bibfnamefont{A.}~\bibnamefont{{Klypin}}},
  \bibinfo{author}{\bibfnamefont{D.~J.} \bibnamefont{{Schlegel}}},
  \bibinfo{author}{\bibfnamefont{S.}~\bibnamefont{{Gottloeber}}},
  \bibinfo{author}{\bibfnamefont{A.~D.} \bibnamefont{{Montero-Dorta}}},
  \bibinfo{author}{\bibfnamefont{M.}~\bibnamefont{{Manera}}},
  \bibinfo{author}{\bibfnamefont{C.~K.} \bibnamefont{{McBride}}},
  \bibinfo{author}{\bibfnamefont{A.~J.} \bibnamefont{{Ross}}},
  \bibnamefont{et~al.}, \bibinfo{journal}{ArXiv e-prints}
  (\bibinfo{year}{2012}), \eprint{1202.6057}.

\bibitem[{\citenamefont{{Albrecht} et~al.}(2006)\citenamefont{{Albrecht},
  {Bernstein}, {Cahn}, {Freedman}, {Hewitt}, {Hu}, {Huth}, {Kamionkowski},
  {Kolb}, {Knox} et~al.}}]{2006astro.ph..9591A}
\bibinfo{author}{\bibfnamefont{A.}~\bibnamefont{{Albrecht}}},
  \bibinfo{author}{\bibfnamefont{G.}~\bibnamefont{{Bernstein}}},
  \bibinfo{author}{\bibfnamefont{R.}~\bibnamefont{{Cahn}}},
  \bibinfo{author}{\bibfnamefont{W.~L.} \bibnamefont{{Freedman}}},
  \bibinfo{author}{\bibfnamefont{J.}~\bibnamefont{{Hewitt}}},
  \bibinfo{author}{\bibfnamefont{W.}~\bibnamefont{{Hu}}},
  \bibinfo{author}{\bibfnamefont{J.}~\bibnamefont{{Huth}}},
  \bibinfo{author}{\bibfnamefont{M.}~\bibnamefont{{Kamionkowski}}},
  \bibinfo{author}{\bibfnamefont{E.~W.} \bibnamefont{{Kolb}}},
  \bibinfo{author}{\bibfnamefont{L.}~\bibnamefont{{Knox}}},
  \bibnamefont{et~al.}, \bibinfo{journal}{ArXiv Astrophysics e-prints}
  (\bibinfo{year}{2006}), \eprint{arXiv:astro-ph/0609591}.

\bibitem[{\citenamefont{{Laureijs} et~al.}(2011)\citenamefont{{Laureijs},
  {Amiaux}, {Arduini}, {Augu{\`e}res}, {Brinchmann}, {Cole}, {Cropper},
  {Dabin}, {Duvet}, {Ealet} et~al.}}]{2011arXiv1110.3193L}
\bibinfo{author}{\bibfnamefont{R.}~\bibnamefont{{Laureijs}}},
  \bibinfo{author}{\bibfnamefont{J.}~\bibnamefont{{Amiaux}}},
  \bibinfo{author}{\bibfnamefont{S.}~\bibnamefont{{Arduini}}},
  \bibinfo{author}{\bibfnamefont{J.~.} \bibnamefont{{Augu{\`e}res}}},
  \bibinfo{author}{\bibfnamefont{J.}~\bibnamefont{{Brinchmann}}},
  \bibinfo{author}{\bibfnamefont{R.}~\bibnamefont{{Cole}}},
  \bibinfo{author}{\bibfnamefont{M.}~\bibnamefont{{Cropper}}},
  \bibinfo{author}{\bibfnamefont{C.}~\bibnamefont{{Dabin}}},
  \bibinfo{author}{\bibfnamefont{L.}~\bibnamefont{{Duvet}}},
  \bibinfo{author}{\bibfnamefont{A.}~\bibnamefont{{Ealet}}},
  \bibnamefont{et~al.}, \bibinfo{journal}{ArXiv e-prints}
  (\bibinfo{year}{2011}), \eprint{1110.3193}.

\bibitem[{\citenamefont{{Bernardeau} et~al.}(2002)\citenamefont{{Bernardeau},
  {Colombi}, {Gazta{\~n}aga}, and {Scoccimarro}}}]{2002PhR...367....1B}
\bibinfo{author}{\bibfnamefont{F.}~\bibnamefont{{Bernardeau}}},
  \bibinfo{author}{\bibfnamefont{S.}~\bibnamefont{{Colombi}}},
  \bibinfo{author}{\bibfnamefont{E.}~\bibnamefont{{Gazta{\~n}aga}}},
  \bibnamefont{and}
  \bibinfo{author}{\bibfnamefont{R.}~\bibnamefont{{Scoccimarro}}},
  \bibinfo{journal}{\physrep} \textbf{\bibinfo{volume}{367}},
  \bibinfo{pages}{1} (\bibinfo{year}{2002}).

\bibitem[{\citenamefont{{Crocce} and
  {Scoccimarro}}(2006{\natexlab{a}})}]{2006PhRvD..73f3519C}
\bibinfo{author}{\bibfnamefont{M.}~\bibnamefont{{Crocce}}} \bibnamefont{and}
  \bibinfo{author}{\bibfnamefont{R.}~\bibnamefont{{Scoccimarro}}},
  \bibinfo{journal}{\prd} \textbf{\bibinfo{volume}{73}},
  \bibinfo{pages}{063519} (\bibinfo{year}{2006}{\natexlab{a}}),
  \eprint{astro-ph/0509418}.

\bibitem[{\citenamefont{{Taruya} and {Hiramatsu}}(2008)}]{2008ApJ...674..617T}
\bibinfo{author}{\bibfnamefont{A.}~\bibnamefont{{Taruya}}} \bibnamefont{and}
  \bibinfo{author}{\bibfnamefont{T.}~\bibnamefont{{Hiramatsu}}},
  \bibinfo{journal}{\apj} \textbf{\bibinfo{volume}{674}}, \bibinfo{pages}{617}
  (\bibinfo{year}{2008}), \eprint{0708.1367}.

\bibitem[{\citenamefont{{Pietroni}}(2008)}]{2008JCAP...10..036P}
\bibinfo{author}{\bibfnamefont{M.}~\bibnamefont{{Pietroni}}},
  \bibinfo{journal}{\jcap} \textbf{\bibinfo{volume}{10}}, \bibinfo{pages}{36}
  (\bibinfo{year}{2008}), \eprint{0806.0971}.

\bibitem[{\citenamefont{{Bernardeau} et~al.}(2008)\citenamefont{{Bernardeau},
  {Crocce}, and {Scoccimarro}}}]{2008PhRvD..78j3521B}
\bibinfo{author}{\bibfnamefont{F.}~\bibnamefont{{Bernardeau}}},
  \bibinfo{author}{\bibfnamefont{M.}~\bibnamefont{{Crocce}}}, \bibnamefont{and}
  \bibinfo{author}{\bibfnamefont{R.}~\bibnamefont{{Scoccimarro}}},
  \bibinfo{journal}{\prd} \textbf{\bibinfo{volume}{78}},
  \bibinfo{pages}{103521} (\bibinfo{year}{2008}), \eprint{0806.2334}.

\bibitem[{\citenamefont{{Pueblas} and
  {Scoccimarro}}(2009)}]{2009PhRvD..80d3504P}
\bibinfo{author}{\bibfnamefont{S.}~\bibnamefont{{Pueblas}}} \bibnamefont{and}
  \bibinfo{author}{\bibfnamefont{R.}~\bibnamefont{{Scoccimarro}}},
  \bibinfo{journal}{\prd} \textbf{\bibinfo{volume}{80}},
  \bibinfo{pages}{043504} (\bibinfo{year}{2009}), \eprint{0809.4606}.

\bibitem[{\citenamefont{{Valageas}}(2010)}]{2010arXiv1009.0106V}
\bibinfo{author}{\bibfnamefont{P.}~\bibnamefont{{Valageas}}},
  \bibinfo{journal}{ArXiv e-prints}  (\bibinfo{year}{2010}),
  \eprint{1009.0106}.

\bibitem[{\citenamefont{{Pietroni} et~al.}(2011)\citenamefont{{Pietroni},
  {Mangano}, {Saviano}, and {Viel}}}]{PieManSav1108}
\bibinfo{author}{\bibfnamefont{M.}~\bibnamefont{{Pietroni}}},
  \bibinfo{author}{\bibfnamefont{G.}~\bibnamefont{{Mangano}}},
  \bibinfo{author}{\bibfnamefont{N.}~\bibnamefont{{Saviano}}},
  \bibnamefont{and} \bibinfo{author}{\bibfnamefont{M.}~\bibnamefont{{Viel}}},
  \bibinfo{journal}{ArXiv e-prints}  (\bibinfo{year}{2011}),
  \eprint{1108.5203}.

\bibitem[{\citenamefont{{Carrasco} et~al.}(2012)\citenamefont{{Carrasco},
  {Hertzberg}, and {Senatore}}}]{2012arXiv1206.2926C}
\bibinfo{author}{\bibfnamefont{J.~J.~M.} \bibnamefont{{Carrasco}}},
  \bibinfo{author}{\bibfnamefont{M.~P.} \bibnamefont{{Hertzberg}}},
  \bibnamefont{and}
  \bibinfo{author}{\bibfnamefont{L.}~\bibnamefont{{Senatore}}},
  \bibinfo{journal}{ArXiv e-prints}  (\bibinfo{year}{2012}),
  \eprint{1206.2926}.

\bibitem[{\citenamefont{{Scoccimarro}}(1998)}]{1998MNRAS.299.1097S}
\bibinfo{author}{\bibfnamefont{R.}~\bibnamefont{{Scoccimarro}}},
  \bibinfo{journal}{\mnras} \textbf{\bibinfo{volume}{299}},
  \bibinfo{pages}{1097} (\bibinfo{year}{1998}),
  \eprint{arXiv:astro-ph/9711187}.

\bibitem[{\citenamefont{{Scoccimarro}}(2001)}]{2001NYASA.927...13S}
\bibinfo{author}{\bibfnamefont{R.}~\bibnamefont{{Scoccimarro}}}, in
  \emph{\bibinfo{booktitle}{The Onset of Nonlinearity in Cosmology}}, edited by
  \bibinfo{editor}{\bibfnamefont{J.~N.} \bibnamefont{{Fry}}},
  \bibinfo{editor}{\bibfnamefont{J.~R.} \bibnamefont{{Buchler}}},
  \bibnamefont{and} \bibinfo{editor}{\bibfnamefont{H.}~\bibnamefont{{Kandrup}}}
  (\bibinfo{year}{2001}), vol. \bibinfo{volume}{927} of
  \emph{\bibinfo{series}{New York Academy Sciences Annals}}, pp.
  \bibinfo{pages}{13--+}.

\bibitem[{\citenamefont{{Crocce} et~al.}(2012)\citenamefont{{Crocce},
  {Scoccimarro}, and {Bernardeau}}}]{2012arXiv1207.1465C}
\bibinfo{author}{\bibfnamefont{M.}~\bibnamefont{{Crocce}}},
  \bibinfo{author}{\bibfnamefont{R.}~\bibnamefont{{Scoccimarro}}},
  \bibnamefont{and}
  \bibinfo{author}{\bibfnamefont{F.}~\bibnamefont{{Bernardeau}}},
  \bibinfo{journal}{ArXiv e-prints}  (\bibinfo{year}{2012}),
  \eprint{1207.1465}.

\bibitem[{\citenamefont{{Audren} and
  {Lesgourgues}}(2011)}]{2011JCAP...10..037A}
\bibinfo{author}{\bibfnamefont{B.}~\bibnamefont{{Audren}}} \bibnamefont{and}
  \bibinfo{author}{\bibfnamefont{J.}~\bibnamefont{{Lesgourgues}}},
  \bibinfo{journal}{\jcap} \textbf{\bibinfo{volume}{10}}, \bibinfo{eid}{037}
  (\bibinfo{year}{2011}), \eprint{1106.2607}.

\bibitem[{\citenamefont{{Crocce} and
  {Scoccimarro}}(2006{\natexlab{b}})}]{2006PhRvD..73f3520C}
\bibinfo{author}{\bibfnamefont{M.}~\bibnamefont{{Crocce}}} \bibnamefont{and}
  \bibinfo{author}{\bibfnamefont{R.}~\bibnamefont{{Scoccimarro}}},
  \bibinfo{journal}{\prd} \textbf{\bibinfo{volume}{73}},
  \bibinfo{pages}{063520} (\bibinfo{year}{2006}{\natexlab{b}}),
  \eprint{astro-ph/0509419}.

\bibitem[{\citenamefont{{Bernardeau} et~al.}(2012)\citenamefont{{Bernardeau},
  {van de Rijt}, and {Vernizzi}}}]{2012PhRvD..85f3509B}
\bibinfo{author}{\bibfnamefont{F.}~\bibnamefont{{Bernardeau}}},
  \bibinfo{author}{\bibfnamefont{N.}~\bibnamefont{{van de Rijt}}},
  \bibnamefont{and}
  \bibinfo{author}{\bibfnamefont{F.}~\bibnamefont{{Vernizzi}}},
  \bibinfo{journal}{\prd} \textbf{\bibinfo{volume}{85}}, \bibinfo{eid}{063509}
  (\bibinfo{year}{2012}), \eprint{1109.3400}.

\bibitem[{\citenamefont{{Bernardeau} et~al.}(2010)\citenamefont{{Bernardeau},
  {Crocce}, and {Sefusatti}}}]{2010PhRvD..82h3507B}
\bibinfo{author}{\bibfnamefont{F.}~\bibnamefont{{Bernardeau}}},
  \bibinfo{author}{\bibfnamefont{M.}~\bibnamefont{{Crocce}}}, \bibnamefont{and}
  \bibinfo{author}{\bibfnamefont{E.}~\bibnamefont{{Sefusatti}}},
  \bibinfo{journal}{\prd} \textbf{\bibinfo{volume}{82}},
  \bibinfo{pages}{083507} (\bibinfo{year}{2010}), \eprint{1006.4656}.

\bibitem[{\citenamefont{{Nishimichi} et~al.}(2009)\citenamefont{{Nishimichi},
  {Shirata}, {Taruya}, {Yahata}, {Saito}, {Suto}, {Takahashi}, {Yoshida},
  {Matsubara}, {Sugiyama} et~al.}}]{2009PASJ...61..321N}
\bibinfo{author}{\bibfnamefont{T.}~\bibnamefont{{Nishimichi}}},
  \bibinfo{author}{\bibfnamefont{A.}~\bibnamefont{{Shirata}}},
  \bibinfo{author}{\bibfnamefont{A.}~\bibnamefont{{Taruya}}},
  \bibinfo{author}{\bibfnamefont{K.}~\bibnamefont{{Yahata}}},
  \bibinfo{author}{\bibfnamefont{S.}~\bibnamefont{{Saito}}},
  \bibinfo{author}{\bibfnamefont{Y.}~\bibnamefont{{Suto}}},
  \bibinfo{author}{\bibfnamefont{R.}~\bibnamefont{{Takahashi}}},
  \bibinfo{author}{\bibfnamefont{N.}~\bibnamefont{{Yoshida}}},
  \bibinfo{author}{\bibfnamefont{T.}~\bibnamefont{{Matsubara}}},
  \bibinfo{author}{\bibfnamefont{N.}~\bibnamefont{{Sugiyama}}},
  \bibnamefont{et~al.}, \bibinfo{journal}{Publ. Astron. Soc. Japan}
  \textbf{\bibinfo{volume}{61}}, \bibinfo{pages}{321} (\bibinfo{year}{2009}),
  \eprint{0810.0813}.

\bibitem[{\citenamefont{{Valageas} and
  {Nishimichi}}(2011)}]{2011A&A...527A..87V}
\bibinfo{author}{\bibfnamefont{P.}~\bibnamefont{{Valageas}}} \bibnamefont{and}
  \bibinfo{author}{\bibfnamefont{T.}~\bibnamefont{{Nishimichi}}},
  \bibinfo{journal}{\aap} \textbf{\bibinfo{volume}{527}}, \bibinfo{eid}{A87}
  (\bibinfo{year}{2011}), \eprint{1009.0597}.

\bibitem[{\citenamefont{{Crocce} et~al.}(2006)\citenamefont{{Crocce},
  {Pueblas}, and {Scoccimarro}}}]{2006MNRAS.373..369C}
\bibinfo{author}{\bibfnamefont{M.}~\bibnamefont{{Crocce}}},
  \bibinfo{author}{\bibfnamefont{S.}~\bibnamefont{{Pueblas}}},
  \bibnamefont{and}
  \bibinfo{author}{\bibfnamefont{R.}~\bibnamefont{{Scoccimarro}}},
  \bibinfo{journal}{\mnras} \textbf{\bibinfo{volume}{373}},
  \bibinfo{pages}{369} (\bibinfo{year}{2006}), \eprint{arXiv:astro-ph/0606505}.

\bibitem[{\citenamefont{Komatsu et~al.}(2009)}]{Komatsu:2008hk}
\bibinfo{author}{\bibfnamefont{E.}~\bibnamefont{Komatsu}} \bibnamefont{et~al.}
  (\bibinfo{collaboration}{WMAP}), \bibinfo{journal}{Astrophys. J. Suppl.}
  \textbf{\bibinfo{volume}{180}}, \bibinfo{pages}{330} (\bibinfo{year}{2009}),
  \eprint{0803.0547}.

\bibitem[{\citenamefont{{Lewis} et~al.}(2000)\citenamefont{{Lewis},
  {Challinor}, and {Lasenby}}}]{2000ApJ...538..473L}
\bibinfo{author}{\bibfnamefont{A.}~\bibnamefont{{Lewis}}},
  \bibinfo{author}{\bibfnamefont{A.}~\bibnamefont{{Challinor}}},
  \bibnamefont{and}
  \bibinfo{author}{\bibfnamefont{A.}~\bibnamefont{{Lasenby}}},
  \bibinfo{journal}{\apj} \textbf{\bibinfo{volume}{538}}, \bibinfo{pages}{473}
  (\bibinfo{year}{2000}), \eprint{arXiv:astro-ph/9911177}.

\bibitem[{\citenamefont{{Taruya} et~al.}(2012)\citenamefont{{Taruya},
  {Bernardeau}, {Nishimichi}, and {Codis}}}]{2012arXiv1208.1191T}
\bibinfo{author}{\bibfnamefont{A.}~\bibnamefont{{Taruya}}},
  \bibinfo{author}{\bibfnamefont{F.}~\bibnamefont{{Bernardeau}}},
  \bibinfo{author}{\bibfnamefont{T.}~\bibnamefont{{Nishimichi}}},
  \bibnamefont{and} \bibinfo{author}{\bibfnamefont{S.}~\bibnamefont{{Codis}}},
  \bibinfo{journal}{ArXiv e-prints}  (\bibinfo{year}{2012}),
  \eprint{1208.1191}.

\bibitem[{\citenamefont{Scoccimarro and Frieman}(1996)}]{Scoccimarro:1996se}
\bibinfo{author}{\bibfnamefont{R.}~\bibnamefont{Scoccimarro}} \bibnamefont{and}
  \bibinfo{author}{\bibfnamefont{J.}~\bibnamefont{Frieman}},
  \bibinfo{journal}{Astrophys. J.} \textbf{\bibinfo{volume}{473}},
  \bibinfo{pages}{620} (\bibinfo{year}{1996}), \eprint{astro-ph/9602070}.

\bibitem[{\citenamefont{{Bernardeau} and {Brax}}(2011)}]{2011JCAP...06..019B}
\bibinfo{author}{\bibfnamefont{F.}~\bibnamefont{{Bernardeau}}}
  \bibnamefont{and} \bibinfo{author}{\bibfnamefont{P.}~\bibnamefont{{Brax}}},
  \bibinfo{journal}{\jcap} \textbf{\bibinfo{volume}{6}}, \bibinfo{pages}{19}
  (\bibinfo{year}{2011}), \eprint{1102.1907}.

\bibitem[{\citenamefont{{Chan} and {Scoccimarro}}(2009)}]{2009PhRvD..80j4005C}
\bibinfo{author}{\bibfnamefont{K.~C.} \bibnamefont{{Chan}}} \bibnamefont{and}
  \bibinfo{author}{\bibfnamefont{R.}~\bibnamefont{{Scoccimarro}}},
  \bibinfo{journal}{\prd} \textbf{\bibinfo{volume}{80}},
  \bibinfo{pages}{104005} (\bibinfo{year}{2009}), \eprint{0906.4548}.

\bibitem[{\citenamefont{{Chuen Chan} et~al.}(2012)\citenamefont{{Chuen Chan},
  {Scoccimarro}, and {Sheth}}}]{2012arXiv1201.3614C}
\bibinfo{author}{\bibfnamefont{K.}~\bibnamefont{{Chuen Chan}}},
  \bibinfo{author}{\bibfnamefont{R.}~\bibnamefont{{Scoccimarro}}},
  \bibnamefont{and} \bibinfo{author}{\bibfnamefont{R.~K.}
  \bibnamefont{{Sheth}}}, \bibinfo{journal}{ArXiv e-prints}
  (\bibinfo{year}{2012}), \eprint{1201.3614}.

\bibitem[{\citenamefont{{Anselmi} et~al.}(2011)\citenamefont{{Anselmi},
  {Matarrese}, and {Pietroni}}}]{2011JCAP...06..015A}
\bibinfo{author}{\bibfnamefont{S.}~\bibnamefont{{Anselmi}}},
  \bibinfo{author}{\bibfnamefont{S.}~\bibnamefont{{Matarrese}}},
  \bibnamefont{and}
  \bibinfo{author}{\bibfnamefont{M.}~\bibnamefont{{Pietroni}}},
  \bibinfo{journal}{\jcap} \textbf{\bibinfo{volume}{6}}, \bibinfo{pages}{15}
  (\bibinfo{year}{2011}), \eprint{1011.4477}.

\bibitem[{\citenamefont{{Sefusatti} and
  {Vernizzi}}(2011)}]{2011JCAP...03..047S}
\bibinfo{author}{\bibfnamefont{E.}~\bibnamefont{{Sefusatti}}} \bibnamefont{and}
  \bibinfo{author}{\bibfnamefont{F.}~\bibnamefont{{Vernizzi}}},
  \bibinfo{journal}{\jcap} \textbf{\bibinfo{volume}{3}}, \bibinfo{pages}{47}
  (\bibinfo{year}{2011}), \eprint{1101.1026}.

\end{thebibliography}

\end{document}